\documentclass[twocolumn]{aastex631}

\usepackage{graphicx}	% Including figure files
\usepackage{amsmath}	% Advanced maths commands
\usepackage{amssymb}	% Extra maths symbols

\usepackage{bm}		% Bold maths symbols, including upright 

\def\lea{\mathrel{<\kern-1.0em\lower0.9ex\hbox{$\sim$}}}
\def\gea{\mathrel{>\kern-1.0em\lower0.9ex\hbox{$\sim$}}}

\usepackage[T1]{fontenc}
\usepackage{ae,aecompl}

\usepackage{newtxtext,newtxmath}

\begin{document}

\title{Clusters, Clumps, Dust, $\&$ Gas (CCDG) in NGC\,1614: Bench-marking Cluster Demographics in Extreme Systems}

\correspondingauthor{Miranda Caputo}
\email{Miranda.Caputo@rockets.utoledo.edu}

\shortauthors{Caputo et al.}
\author[0000-0002-2957-3924]{Miranda Caputo}
\affiliation{Ritter Astrophysical  Research Center, University of Toledo, Toledo, OH 43606, USA}
\author[0000-0003-0085-4623]{Rupali Chandar}
\affiliation{Ritter Astrophysical  Research Center, University of Toledo, Toledo, OH 43606, USA}
\author[0000-0001-7413-7534]{Angus Mok} \affiliation{OCAD University, Toronto, Ontario, M5T 1W1, Canada}
\author{Sean Linden}
\affiliation{Steward Observatory, University of Arizona, 933N Cherry Avenue, Tucson, AZ 85721, USA}
\author[0000-0002-5728-1427]{Paul Goudfrooij}
\affiliation{Space Telescope Science Institute, 3700 San Martin Drive, Baltimore, MD, 21218, USA}
\author{Bradley C. Whitmore}
\affiliation{Space Telescope Science Institute, 3700 San Martin}

% These dates will be filled out by the publisher
%\date{Last updated 2015 May 22; in original form 2013 September 5}

% Enter the current year, for the copyright statements etc.

% Don't change these lines
%\begin{document}
%\label{firstpage}
%\maketitle

\begin{abstract}

Observations of young star clusters in a variety of galaxies have been used to constrain basic properties related to star-formation, such as the fraction of stars found in clusters ($\Gamma$) and the shape of the cluster mass function. 
However, the results can depend heavily on the reliability of the cluster age-dating process and other assumptions.
One of the biggest challenges for successful age-dating lies in breaking the age-reddening degeneracy, where older, dust-free clusters and young, reddened clusters can have similar broad-band colors. 
While this degeneracy affects cluster populations in all galaxies, it is particularly challenging in dusty, extreme star-forming environments systems.
We study the cluster demographics in the luminous infrared galaxy NGC\,1614 using Hubble imaging taken in 8 optical-near infrared passbands. 
For age-dating, we adopt a spectral energy distribution fitting process that limits the maximum allowed reddening by region, and includes H$\alpha$ photometry directly.
We find that without these assumptions, essentially all clusters in the dust-free UV-bright arm which should have ages $\approx50-250$~Myr are incorrectly assigned ages younger than 10~Myr.
We find this method greatly reduces the number of clusters in the youngest ($\tau$ $<$ 10~Myrs) age bin and shows a fairly uniform distribution of massive clusters, the most massive being $\approx \mbox{few} \times 10^7~\mathrm{M}_{\odot}$.
A maximum likelihood fit shows that the cluster mass function is well fitted by a power-law with an index $\sim$-1.8, with no statistically significant high-mass cutoff. 
We calculate the fraction of stars born in clusters to be $\Gamma_{1-10}$ = 22.4$\%$ $\pm$ 5.7$\%$.  The fraction of stars in clusters decreases quickly over time, with $\Gamma_{10-100} = 4.5\pm 1.1$\% and $\Gamma_{100-400}=1.7\pm 0.4$\%, suggesting that clusters dissolve rapidly over the first $\sim0.5$~Gyr.  The decreasing fraction of stars in clusters is consistent with the declining shape observed for the cluster age distribution.

\end{abstract}


%%%%%%%%%%%%%%%%% BODY OF PAPER %%%%%%%%%%%%%%%%%%
\section{Introduction}

Extremely massive star clusters ($\sim 10^6$-$10^8$~M$_\odot$), which may be young analogs of ancient globular clusters, are forming in extreme star-forming galaxies in the nearby ($<$ 100~Mpc) universe \citep{Portegies10,Linden17,Adamo20}.
The star formation in many of these systems is triggered by minor or major merging events between gas-rich galaxies, which can lead to an infrared-luminous phase.
Nearby merging systems give important insight into the star and cluster formation processes that operated near cosmic noon, z$\sim$1.5-3, when merging was much more common than in the modern universe \citep[e.g.,][]{Bridge07,Carpineti15,Conselice14}.

Studying intensely star-forming galaxies, where the star formation rate is $>$ 2$\sigma$ above the star-formating main sequence (e.g., \citealt{Speagle14}), allows us to probe the high end of fundamental relationships that may depend on star formation rate (SFR in M$_\odot$~yr$^{-1}$) or SFR density ($\Sigma_{\rm SFR}$ in M$_\odot$~yr$^{-1}$~kpc$^{-2}$). One example is the cluster mass function, which encodes important information about the formation and evolution of cluster populations in galaxies.
Most current simulations either find or assume the cluster mass function has a Schechter-like distribution, with an upper mass cutoff that increases with $\Sigma_{\rm SFR}$ of the galaxy \citep{Li17,Li18}.
Observational works find somewhat mixed results for the shape of the cluster mass function, with some favoring a Schechter-like distribution \citep{Portegies10,Bastian12,Johnson17,Messa18b,Adamo20}, while others find that a single power-law fits the observations well \citep{whitmore10,cook19,Adamo20,Chandar14,Chandar16,Chandar23b}.  

Another parameter that may vary with SFR and $\Sigma_{\rm SFR}$ is the fraction of stars found in clusters, $\Gamma$.
We assume that the fraction of stars found in 1-10~Myr clusters is a good proxy for the fraction {\em born} in clusters.
A number of observational works have suggested that the fraction of stars found in clusters with ages between 1 and 10~Myr increases with $\Sigma_{\rm SFR}$ \citep{Goddard10,Johnson16,Adamo15,Adamo20} from just a few percent to nearly 100\%.
%, since we expect cluster disruption over the first 10~Myr to be low \citep{fall09}. }
However, this result has been called into question due to clusters of different ages being used to estimate $\Gamma$ in galaxies with low vs. high $\Sigma_{\rm SFR}$ \citep{Chandar17}.  Newer works suggest that potential issues with cluster age-dating may also have affected previous results for $\Gamma$ \citep{Chandar23b}.

One key challenge in determining the shape of the cluster mass function and in calculating $\Gamma$ in nearby star forming galaxies arises from the difficulty of breaking the age-reddening degeneracy of clusters in these dusty systems based on broad-band measurements alone.
One improvement is to include H$\alpha$ in the age-dating procedure, as \cite{whitmore20} did to improve cluster ages in the dwarf starburst galaxy NGC 4449.
Recently, \cite{Whitmore23b} found that ancient globular clusters in nearby spiral galaxies were being assigned ages that are too young by factors of 10 to 1000, but that including H$\alpha$ or CO information significantly improved the age-dating results for globular clusters.
\cite{Chandar23b} and \cite{Chandar23a} found that including H$\alpha$ measurements combined with limiting the maximum allowed E(B-V) in the SED fitting improved the resulting ages, and in particular reduced the number of clusters incorrectly dated to ages younger than 10~Myr.

In this work we will focus on the cluster population in NGC\,1614, which was observed as a part of the Clusters, Clumps, Dust, and Gas (CCDG) multi-band HST imaging survey of 13 extreme star-forming galaxies.
NGC~1614 (Figure \ref{FIG:RGB_IVB}) is a minor merger between a spiral and a dwarf galaxy, is located at $D \sim$\,69.7 Mpc, and is a luminous infrared galaxy (LIRG) (L$_{IR}$/L$_\odot$ = 11.6). 
NGC\,1614 has a variety of environments including: tidal tails, a star-forming dusty arm in the west, a nearly dust-free UV-bright arm in the east (similar to UV-bright clumps observed in galaxies at z$\sim1-2$), and a dust-enshrouded central region.
AGN activity has not been detected from the center of NGC\,1614 \citep{Herrero-Illana17}. 
Due to these different environments which have different amounts of dust,  NGC\,1614 it is a good target for testing assumptions related to reddening (E(B-V)) during the age-dating process. 

The remainder of this paper is organized as follows: 
Section 2 looks at the properties of NGC\,1614 and presents the HST data with source detection in the optical and a few clusters discovered in the NIR imaging.
Section 3 focuses on cluster analysis and methods to handle the age-reddening degeneracy.
Section 4 presents the results found for the cluster age and mass estimates, cluster mass functions, age distributions, and $\Gamma$.
We discuss the physical implications in Section 5.
Section 6 gives a summary of our conclusion, and in an Appendix we discuss uncertainties in the assumptions used to calculate $\Gamma$ and implications for published results in NGC~1614 and other galaxies.

%%%%%%%%%%%%%%%%%%%%%%%%%%%%%%%%%%%%%%%%%%%%%%%%%%
\section{NGC\,1614: Properties, Observations, and Cluster Catalog}

\subsection{Star Formation Rate}
\label{SEC:sfr}

In this paper, we aim to understand the relationship between star and star cluster formation in NGC\,1614. In order to do this, we establish the star formation rate (SFR) and SFR per unit area ($\Sigma_{\rm SFR}$) that we will use for NGC\,1614 in this section. 

NGC\,1614 has had several SFR estimates over the past decade using a variety of tracers, including hydrogen recombination lines tracing very recent ($\tau$ $\le$ 10 Myr) star formation and infrared luminosity which includes emission from older ($\tau$ $\approx$ few$\times$ 100~Myr) populations \citep{Murphy12}. 

Some previously estimated SFRs from hydrogen lines for NGC\,1614 include: 
(1) 27.4 M$_{\odot}$ yr$^{-1}$  based on HST continuum subtracted H$\alpha$ emission corrected for the median extinction found from young clusters \citep{Adamo20}; 
(2) 49.6 M$_{\odot}$ yr$^{-1}$ from HST Pa$\beta$ emission after 
applying a dust correction using the Balmer-to-Paschen decrement (H$\alpha$/Pa$\beta$) \citep{Gimenez-Arteaga22} ;
(3) 74.7 M$_{\odot}$ yr$^{-1}$ from dust-extinction corrected Pa$\alpha$ emission \citep{Tateuchi15}.

IR emission tends to trace stellar populations that have ages 
up to $\sim$500\,Myr \citep{Kennicutt12}. For NGC\,1614, IR based SFR estimates include:
(1)  41.8 M$_{\odot}$ yr$^{-1}$ based on F110W luminosity from HST to 95.7 M$_{\odot}$ yr$^{-1}$ when 24 micron flux is included \citep{Gimenez-Arteaga22};
(2) 49.0 M$_{\odot}$ yr$^{-1}$ based on 8-1000 micron emission \citep{Tateuchi15};
(3) and 51.3 M$_{\odot}$ yr$^{-1}$ from IRAS IR data \citep{Linden17}.

While there is a range in published SFR estimates for both the hydrogen line and IR emission, there is reasonable agreement between estimates which use these two tracers, which give an average of $\sim$51 and $\sim$59 M$_{\odot}$ yr$^{-1}$ for hydrogen lines and IR emission, respectively. This suggests that the SFR has been fairly constant over at least the last $\sim$100 Myr, and likely $\approx0.5$~Gyr.
In this work, we adopt a SFR for NGC\,1614 of 49.6 $\ M_{\odot} yr^{-1}$ from \cite{Gimenez-Arteaga22} based on hydrogen recombination lines, since they trace the most recent ($<10$~Myr) star formation. This rate is quite similar to the mean values found between the different published results for hydrogen lines and from IR emission. We adopt an uncertainty of $25$\% following \cite{Cook23} and \cite{Chandar23b}.

We estimate the area of NGC\,1614 to be $\sim$200~kpc$^{-2}$. This region includes essentially all of the clusters in our sample plus parts of the tidal tails to the south, east, and north. 
With our adopted SFR of $49.6~M_{\odot} \mbox{yr}^{-1}$ and area of 200~kpc$^2$, $\Sigma_{\rm SFR}=$ 0.25~$\ M_{\odot}~{\rm yr}^{-1}$~kpc$^{-2}$.
\cite{Adamo20} adopted a smaller area of 81~kpc$^{-2}$, but excluded the diffuse emission towards the north and parts of tidal features which are included in our work.
With their SFR estimate of 27.4$\ M_{\odot}\mbox{yr}^{-1}$, \cite{Adamo20} assumed $\Sigma_{\rm SFR}=$ 0.34~$\ M_{\odot}~{\rm yr}^{-1}$~kpc$^{-2}$.

\subsection{Optical Source Detection \& Photometry}

\label{SEC:data}

We use HST images taken in the NUV (F275W), U (F336W), B (F438W), V (F555W), H$\alpha$ (F665N), I (F814W), Paschen$\beta$ (F130N), and H (F160W) passbands.
The observations used in this work are a mix of new WFC3 and archival ACS (B and I)images, with the new data taken as part of the Clusters, Clumps, Dust, and Gas survey of extreme galaxies (CCDG) program (GO-15649; PI: Chandar) with the WFC3 camera.

Each individual exposure is processed though the standard Pyraf/STSDAS CALACS or CALWFC3 software before alignment and drizzling to a common grid using DRIZZLEPAC, creating a single, sky-subtracted image for each filter. 
Gaia DR2 sources \citep{Gaia18} are used to astrometrically correct the V-band image, which is used as a reference for all other filters. 
FITS file outputs are in units of electrons per second, and oriented with north up and east to the left.

For source detection, we use the DAOFIND detection algorithm in IRAF with a 3$\sigma$ detection limit on the V-band image. At a distance of $\sim$69.7~Mpc, clusters in NGC\,1614 appear as point-like sources in the HST data. In order to minimize effects from crowding and scatter in the measured colors, aperture photometry with a 2 pixel radius and background annuli between 7 and 9 pixels was performed on all sources in each filter.
Background levels were determined as the median flux value after sigma clipping.
Aperture corrections, were determined from encircled energies derived from point sources in each filter, of  1.21~mag (F275W), 1.10~mag (F336W), 0.890~mag (F435W), 0.85~mag (F555W), 1.02~mag (F814W), and 0.76~mag (F665N).
Magnitudes are converted to the VEGAmag system.
We do not include background subtraction in the photometry performed on the non- continuum-subtracted H$\alpha$ image, since the warm, ionized gas can have a different morphology than the starlight, and is frequently in shells and rings around young clusters. 
We select clusters to have a measured V-band magnitude brighter than $\sim$26~mag to limit photometric errors particularly in the bluer filters, and to have a concentration index (magnitude difference measured in apertures of 0.5 and 2~pixels)
between 1.1 and 2.0~mag to eliminate extended background sources since clusters are expected to be point sources.
Additional details on reduction and photometry are given in a paper presenting the CCDG survey (Chandar et al., in prep).

\begin{figure*}
    \centering
    \includegraphics[width=0.9\textwidth]{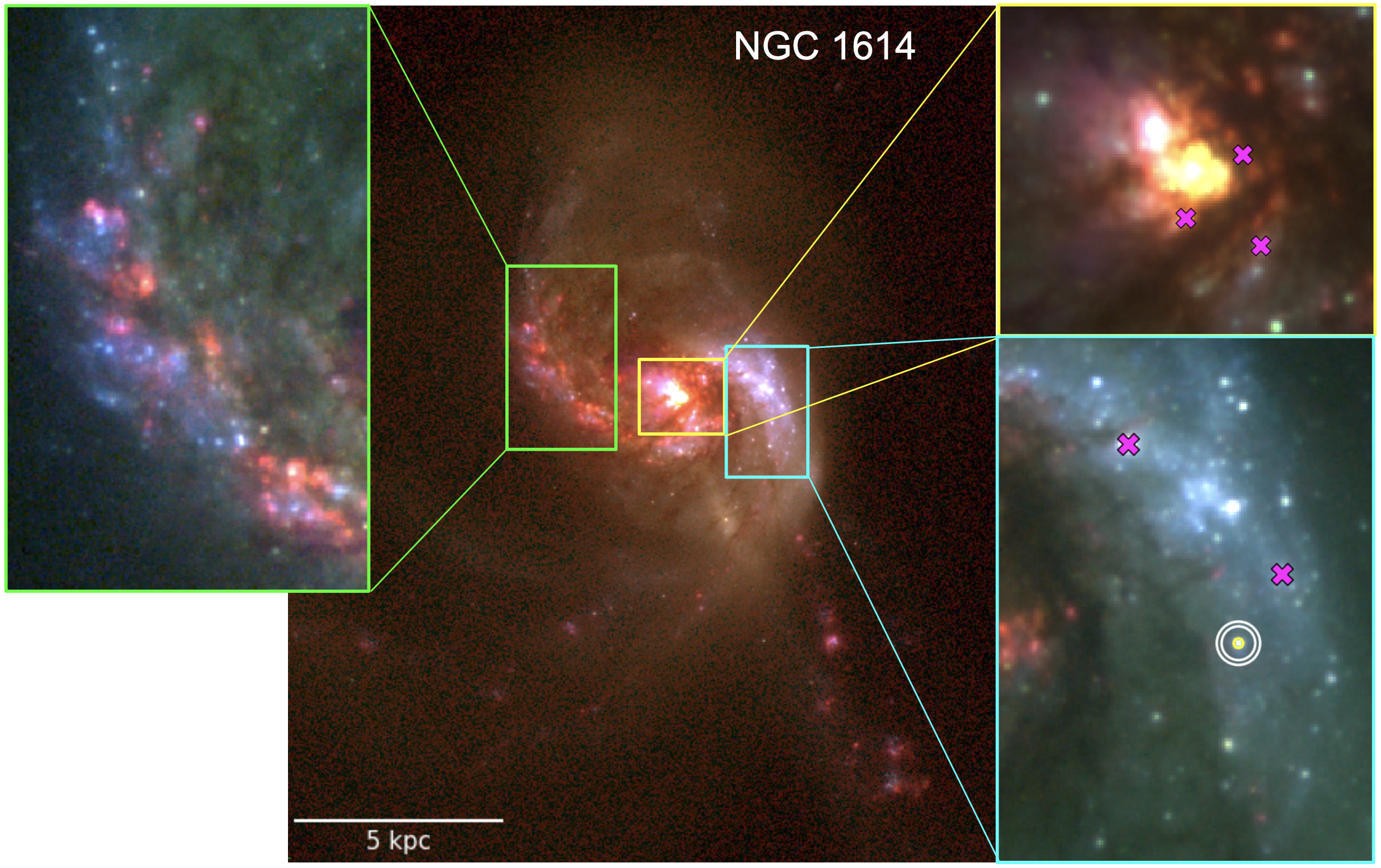}
    \caption{HST image of NGC\,1614 with B, V, and H$_{\alpha}$ represented by blue, green, and red respectively. A variety of environments within this galaxy are highlighted, with the western UV-bright arm, eastern star forming arm, and central region enlarged as shown. The yellow circle represents an aperture with a 2 pixel radius. The white circles show the 7 and 9 pixel radii used to determine the background level. 
    The locations of the five most massive clusters younger than 400\,Myr are shown as the magenta crosses in the two right panels.}
    \label{FIG:RGB_IVB}
\end{figure*}

A color image of NGC\,1614 in the B, V, H$\alpha$ filters is shown in Figure \ref{FIG:RGB_IVB}. Three regions of interest are highlighted: the star-forming eastern arm, the dusty central region, and the dust and H$\alpha$ free UV-bright arm. A 2 pixel radius  yellow circle and annuli of 7 and 9 pixels white circles are shown on a cluster in the UV-bright arm as an example of the apertures used to perform photometry.

\subsection{IR-only Detected sources}
\label{sec:missingIR}
\begin{figure*}
    \centering
    \includegraphics[width=0.9\textwidth]{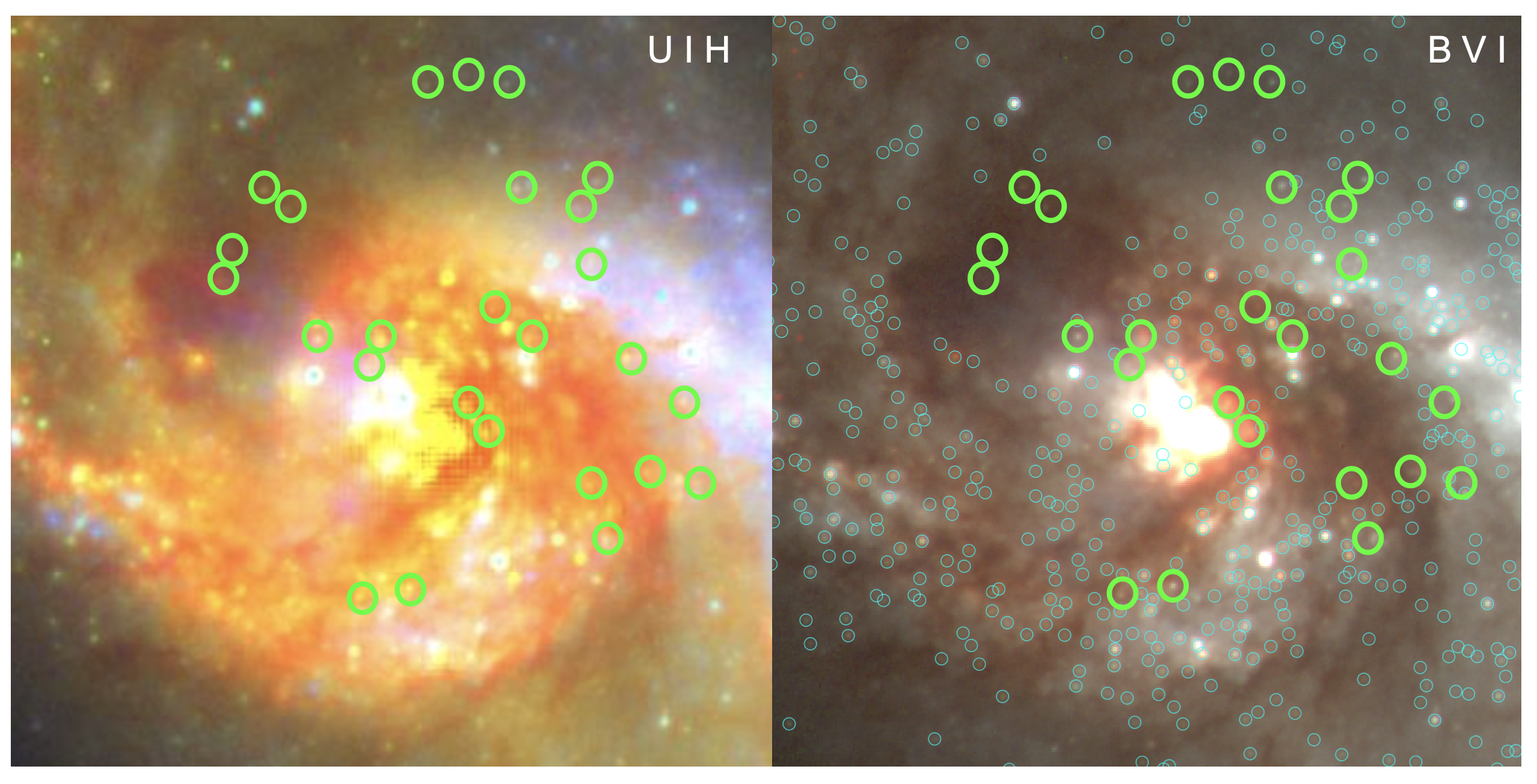}
    \caption{{\bf Left:} U, I, and H-Band HST color image of NGC\,1614 with new sources visible in the near-infrared (H-band) but not in the optical (V-band) circled in lime green. These were identified by eye. {\bf Right:} B, V, I HST color image with optically detected clusters circled in cyan and new NIR-only clusters circled in lime.} 
    \label{FIG:MissingIR}
\end{figure*}
Because NGC~1614 is a luminous infrared galaxy with significant amounts of dust, we use the Pa$\beta$ and H-band infrared images to search for and estimate the number of optically-obscured clusters.
We do not however, include Pa$\beta$ or H-band measurements in our age-dating due to the lower resolution of the IR camera compared with UVIS.

Using both the H-band (F160W) and Pa$\beta$ (F130N) along with a V-band image, we visually searched through the galaxy for any compact, point-like sources that appear in the two NIR filters but are not observed in the V-band.
Figure \ref{FIG:MissingIR} shows a NIR U, I, and H-band color image of NGC\,1614 with each of the clusters identified in the NIR circled in cyan. 
We identify 27 ($\sim2.4$\% of the sample) new sources in the H-band images that were not detected in V-band, although all but 6 ($\sim$0.5$\%$) show at least some faint emission in the I-band. All of the NIR sources are found in dusty areas and a couple in somewhat isolated locations.

We will not include these 27 clusters in the rest of this work as we are unable to age-date them. We are not able to determine if these clusters are younger than 10 million years or not from the F130N Pa$\beta$ filter due to its poorer resolution and overlap with the H-band in the bright and extended emission seen throughout these dusty regions, but will present follow-up results from an accepted JWST proposal (GO-6035; PI: Caputo) in the future.

%%%%%%%%%%%%%%%%%%%%%%%%%%%%%%%%%%%%%%%%%%%%%%%%%%
\section{Cluster Analysis}
\label{SEC:clusters}

In this section, we will estimate the age and mass for each cluster in our sample by comparing their measured magnitudes with predictions from stellar evolutionary models via SED fitting. 
One of the biggest challenges for these estimates lies in navigating the age-reddening degeneracy \citep[e.g.,][]{Whitmore23a}. This is where older clusters --- those that are intrinsically red with little-to-no extinction --- have similar broad-band colors as clusters that are moderately extinguished and intermediate in age or, more likely, young and highly extinguished. 
We use two strategies to resolve the age-reddening degeneracy: (1) use available information to constrain the maximum E(B-V) within sub-regions of the galaxy (see \cite{Chandar23a}), and (2) include H$\alpha$ directly in the SED fits \citep{whitmore10,whitmore20,Chandar23b, Chandar23a}.

\subsection{Color-Color Diagrams and Training Sets}
\label{sec:cctraining}

\begin{figure*}
    \centering
    \includegraphics[width=0.90\textwidth]{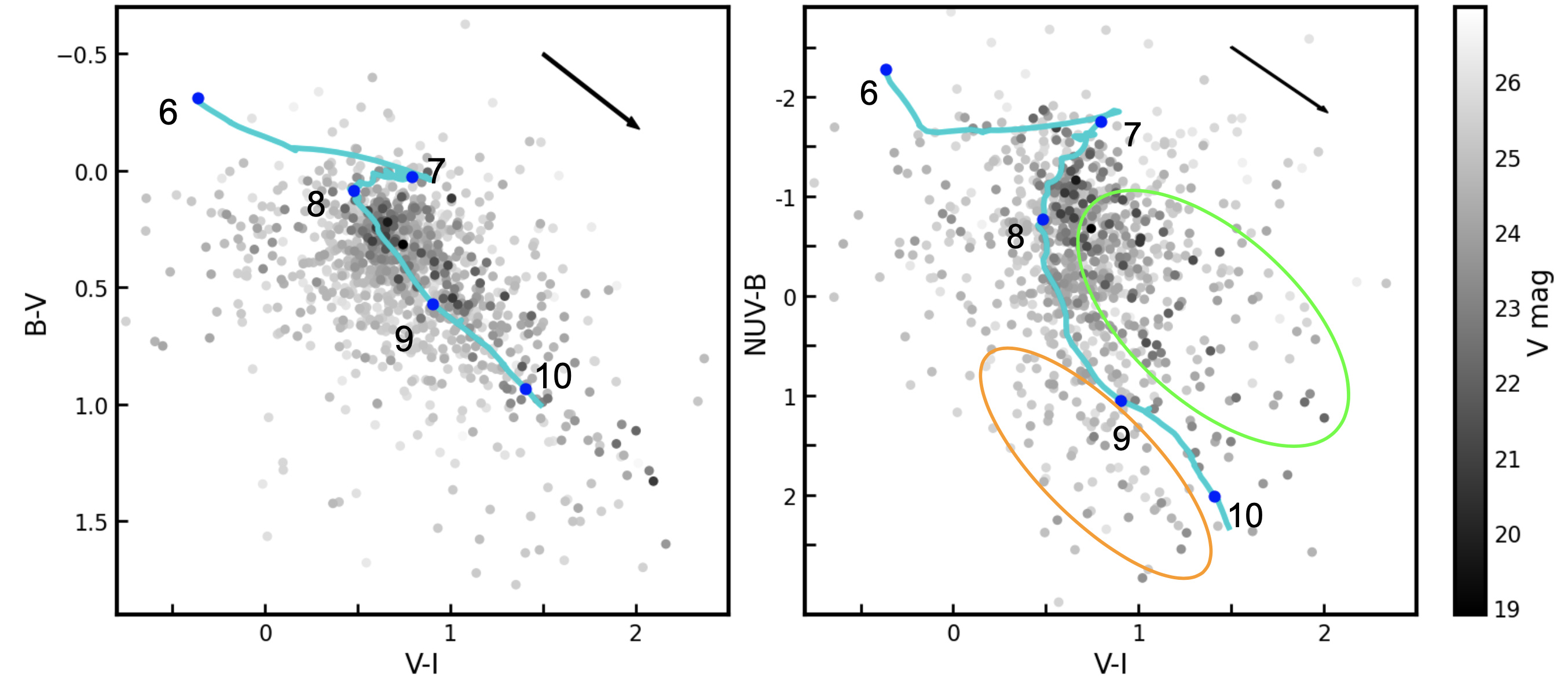}
    \caption{{\bf Left:} Color-color diagrams for clusters in NGC\,1614 with B-V vs V-I shown on the {\bf Left} and NUV-B vs V-I shown on the {\bf Right}. The grayscale showing the V-band magnitude is to the right of the diagrams. Predictions from solar-metallicity Bruzual \& Charlot (2003) tracks, cyan, for cluster evolution are shown in blue  for log~$(\tau/\mbox{yr})=$ 6, 7, 8, 9, and 10. 
    A reddening vector with A$_\mathrm{V}$ = 1.0 is shown in the top right of each panel.
    \label{FIG:CC}}
\end{figure*}

In Figure~\ref{FIG:CC}, we show color-color diagrams with measured magnitudes for the clusters in NGC\,1614.
These diagrams include predictions from solar metallicity \cite{Bruzual03} (hereafter BC03) evolutionary models for cluster colors (plotted in cyan) starting from 1 Myr in the top left to 10+ Gyr in the bottom right, with each factor of 10 in age marked as blue points.  
Clusters are plotted with colors that scale with their V-band magnitude, with the brightest clusters shown as the darkest points.
The direction and amount that reddening will shift measured cluster colors by an A$_\mathrm{V}$ = 1.0~mag is represented by the arrow in the top-right corner of the color-color diagrams, assuming a Milky Way-like reddening law \citep{Fitzpatrick99}.
Our cluster colors are not corrected for foreground extinction since it is fairly small ($\sim$0.4 mag in V-band \citep{Schlafly11}) compared with many clusters in this galaxy.

We find that the brightest clusters in NCG 1614 have a range in reddening. 
While many clusters closely follow the model track (Av $\lea$ 0.3~mag) indicating they experience little-to-no reddening in NGC~1614,   
a subset of bright clusters falls rightward of the model in the direction expected from reddening (Figure \ref{FIG:CC} circled in lime).
A visual examination of this subset shows that most have H$\alpha$ emission, and are therefore quite young ($\le 6$~Myr), recently formed clusters still (partially) embedded in and reddened by their natal gas and dust.
There is another subset of clusters (circled in orange) --- most clearly seen in the NUV-B vs V-I colors --- that fall below the models. These clusters are intermediate in age ($\approx \mbox{few}\times100$~Myr) with a range of reddening much like the H$\alpha$-emitting young clusters. A similar population of reddened, intermediate-age clusters exists in the dusty spiral galaxy NGC~1365 \citep{Whitmore23a}. 

\begin{figure}
    \centering
    \includegraphics[width=0.40\textwidth]{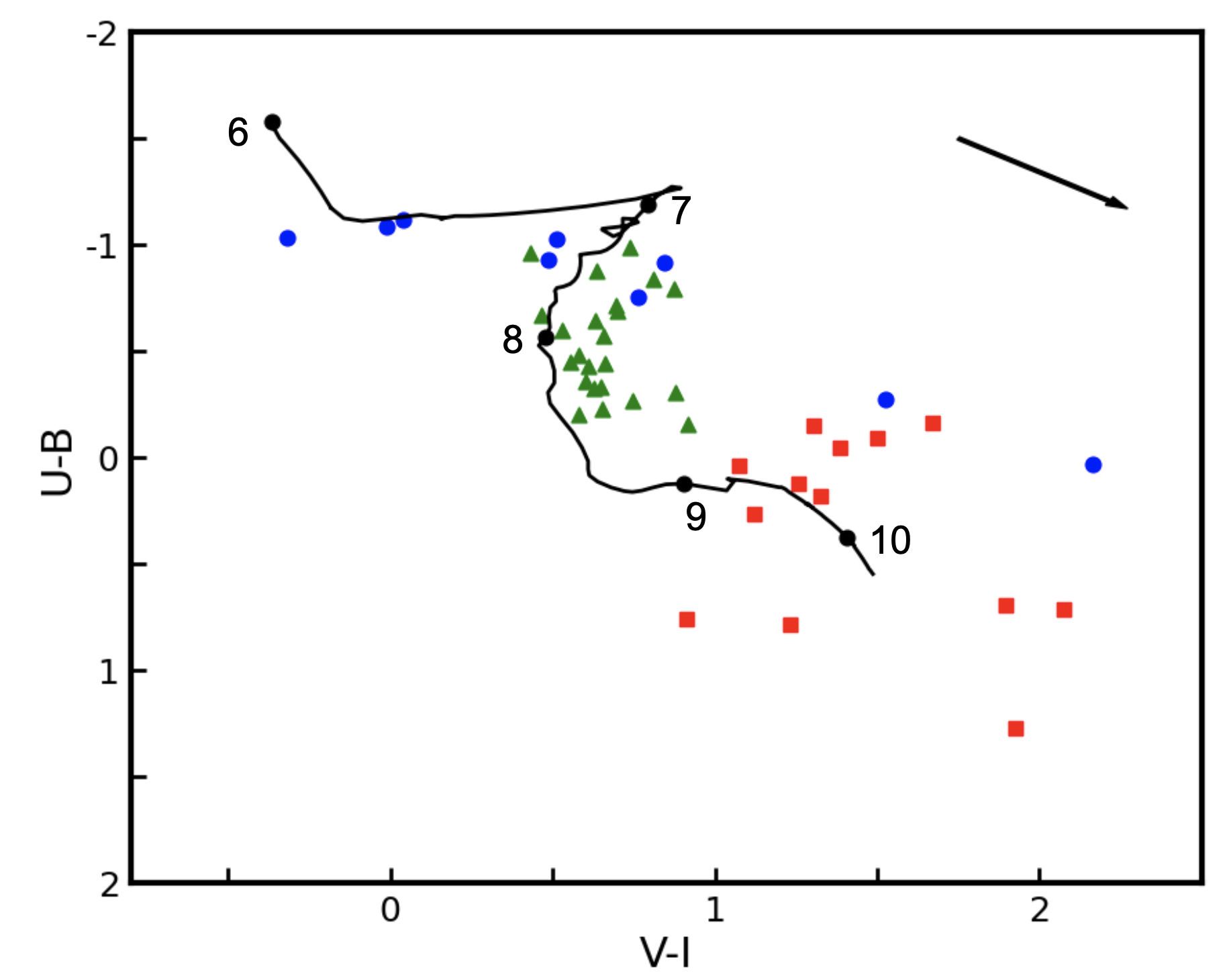}
    \caption{U-B vs V-I color-color diagram with BC03 model in black, H$\alpha$ training set clusters shown as blue circles, UV-arm training set in green triangles, and old cluster training set in red squares. A reddening vector with an A$_\mathrm{V}$ = 1.0 is shown in the top right.}
    \label{FIG:train}
\end{figure}

To establish an appropriate maximum E(B-V) to adopt during the SED fitting procedure (Section \ref{sec:agedating}), we visually select the three training sets seen in Figure \ref{FIG:train}. These sets are comprised of clusters with different ages, amounts of reddening, and magnitudes that are easily visually categorized into the following:

\begin{itemize}
    \item Very bright, young ($\le 6$~Myr) clusters, selected to have strong H$\alpha$ emission and a range of broad-band colors indicating different amounts of reddening
    \item Intermediate age ($\approx 100$~Myr) clusters in the nearly dust-free, western UV-bright arm with a range of V-band magnitudes
    \item Older clusters which appear red and show no obvious dust surrounding them in the images.   
\end{itemize}

In Figure \ref{FIG:train}, young, strong H$\alpha$-emitting clusters (blue) have a large range of colors that fall off the models in the direction expected from reddening. We estimate the reddening of each cluster by comparing their measured colors with those predicted for a 3~Myr old cluster. 
{\em We find that the maximum reddening experienced by clusters varies from one region to another within the galaxy.}  In the next subsection we will use properties of the H$\alpha$-emitting clusters as a prior to set the maximum E(B-V) in different regions within NGC\,1614.

The UV-bright arm provides an important, relatively dust-free laboratory to test parameters for our SED fitting. The clusters in this region follow the predicted model colors well (within Av $\sim$0.3 mag), as seen by the green points in Figure \ref{FIG:train}, indicating they all have similar intermediate ages (but a range of V-magnitudes).  
Since there is essentially no H$\alpha$ emission in this region, clusters in the western UV-bright arm to the west cannot be very young ($\tau \le$ 6~Myr) and reddened. This, paired with their locations on the color-color diagram and lack of dust in the region (see Figure \ref{FIG:CC}) are strong evidence that their ages are on the order $\approx 50-200$~Myr.
We find that restricting the maximum allowed E(B-V) to low values (e.g., 0.1~mag) is important to correctly age-date these intermediate age clusters. 

Finally, a set of clusters that are very red (B-V $\ge$ 0.6 and V-I $\ge$ 0.8), bright (V-mag $\le$ 24), and in relatively dust free areas are identified and plotted in red in Figure \ref{FIG:train}. These clusters are expected to be older (log(age) $\ge$ 8.3), and we can restrict the maximum E(B-V) to ensure their estimated ages reflect this.

%--------------------------------

\begin{figure*}
    \centering
    \includegraphics[width=0.90\textwidth]{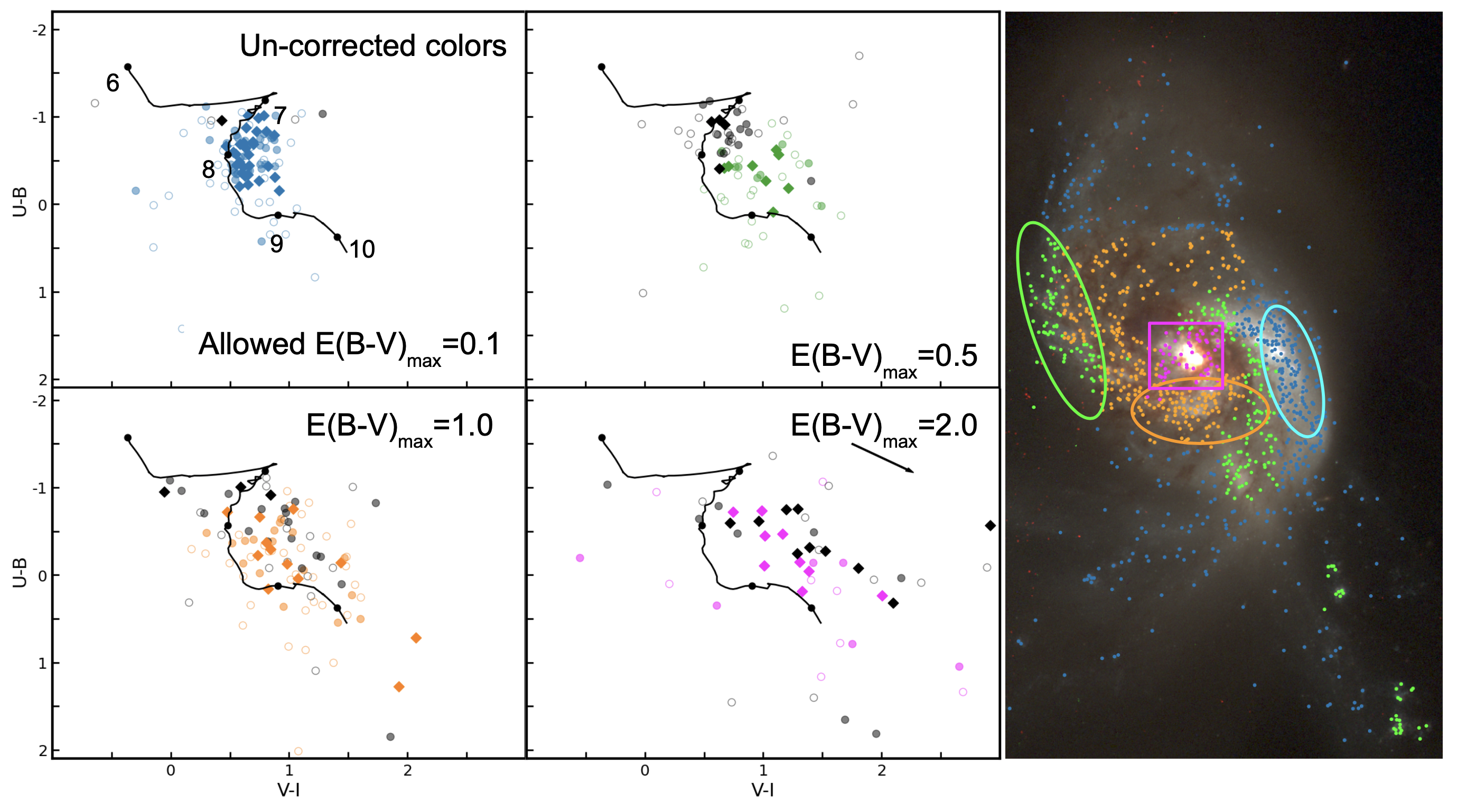}
    \caption{{\bf Right:} B, V, I image of NGC\,1614 with detected clusters overlaid; the color-coding shows the maximum E(B-V) allowed during the age-dating fits. {\bf Left:} Un-corrected U-B vs V-I color-color diagrams of 4 example regions with assigned maximum allowed E(B-V) of 0.1~mag (top-left), 0.5~mag (top-right), 1.0~mag (bottom-left) and 2.0 (bottom-right). A reddening vector of A$_\mathrm{V}$ = 1.0 is shown in the bottom right image.
    Diamonds indicate clusters with V-mag <= 23.5, filled circles represent 23.5 <= V-mag <24.5 and open circles represent V-mag >= 24.5.
    Data points outlined and filled in black indicate the cluster has H$\alpha$ emission and is younger than $\approx6$~Myr.
    These regions illustrate the different amounts of reddening due to dust found for clusters in different regions within NGC 1614. 
    }
    \label{FIG:reddening}
\end{figure*}
%---------------------------------

%--------------------------------

\subsection{Constraints on Reddening By Region}
\label{sec:reddenings}

NGC\,1614 has a large range in reddening as seen by both visual inspection of the galaxy and the colors of young clusters.
We develop a method to establish priors by determining the maximum to E(B-V) value to adopt in the SED fitting, which we allow to vary by region within the galaxy. 
Regions are determined by visual inspection and appear to have similar amounts of dust affecting the cluster colors.
We want the maximum allowed E(B-V) to be 
sufficiently high that reddened, young H$\alpha$ emitting clusters, which are younger than $\sim$6~Myr, are correctly age-dated, but not so high that older, gas-free clusters ($\ge$ 100~Myr) are erroneously fit by young ($\sim$ 10~Myr) ages and high reddening. In regions where the cluster colors hug the models closely and there is no indication of dust, low maximum E(B-V) values are adopted. 

NGC\,1614 is segmented into regions, based on the color-color diagrams of H$\alpha$ emitting clusters, representing one of four maximum allowed E(B-V) values: 0.1, 0.5, 1.0, and 2.0~mag.

Figure~\ref{FIG:reddening} shows the distribution of the clusters by reddening group on a color image to the right, with blue, green, orange, and magenta representing maximum E(B-V) values of 0.1, 0.5, 1.0, and 2.0~mag respectively. 
One subset of each of the four reddening groups is circled on the image to the right, and also has their clusters represented in the U-B vs V-I color-color diagrams in the left panels, with the maximum E(B-V) indicated. 
The progression in reddening is most obviously demonstrated by the young clusters with H$\alpha$ emission, plotted in black, falling further off the models in the direction of the reddening arrow as one moves to dustier regions within the galaxy, as indicated by the increasing E(B-V)${\rm max}$ values.

The region shown with the lowest maximum E(B-V) is the UV-bright arm, shown as the blue points in the top-left panel of Figure \ref{FIG:reddening}.
The cluster colors within this region are very close to the model track (within Av $\lea$0.3 mag) with little evidence for reddening, and thus represent a relatively dust-free region.
In addition, the clusters in this region should be older than 10\,Myr, as there is no detected H$\alpha$ emission and the colors suggest cluster ages closer to $\sim$100\,Myr throughout the region.
If we allow a maximum E(B-V) of $>$0.1,
age estimates for the clusters in this region start piling up around log($\tau/\mbox{yr}$)= 6.8, an age where no H$\alpha$ emission is predicted, but which is erroneously young for these clusters. 
Restricting the maximum E(B-V) to 0.1 for these relatively dust-free clusters allows for accurate age estimates.
We therefore adopt a maximum E(B-V) = 0.1~mag during SED fitting for clusters in the UV-bright arm and others shown in blue across the galaxy.

A portion of the eastern, star-forming arm is shown in green and represents regions with modest reddening. 
Clusters in this region begin to fall further to the right of the models than seen in regions like the UV-bright arm. 
There are a number of somewhat reddened H$\alpha$ emitting clusters to the right of the model. 
These clusters need a somewhat higher E(B-V) in order to be correctly fit to a young age, but do not require more than E(B-V) = 0.5~mag.

The region to the south of the center of NGC\,1614, plotted in the lower-left panel in orange, illustrates a moderately dusty region in the galaxy. 
H$\alpha$ emitting clusters have colors that fall further still along the reddening vector, with some having colors similar to those expected for ancient globular clusters.  
Note however, that we expect to detect very few globular clusters at the distance of NGC~1614. 
This region is given a maximum E(B-V) = 1.0~magto allow the H$\alpha$ emitting clusters to be fit by a young age.

Finally, the central region of NGC\,1614, plotted in magenta, is given the largest maximum E(B-V) = 2.0~mag because of the large amount of dust and highly reddened young clusters in this region, a few of which required an E(B-V) $\ge$ 1.5 to be correctly age-dated. 
Hardly any clusters in this region fall on the BC03 model. 
This is also the only region in NGC 1614 given a maximum E(B-V) $>$ 1.0~mag.
Comparing results for this region when we adopt a maximum E(B-V) of 1.5~mag vs 2.0~mag, only a few clusters, confirmed by eye to have H$\alpha$ emission, correctly move to ages less than 6~Myr with a maximum E(B-V) of 2.0~mag. All other clusters have the same best-fit ages in both cases.

A total of 525 clusters (46.4$\%$ of the total sample), are given the lowest maximum E(B-V) value of 0.1~mag. 
These clusters show little-to-no deviation from the models in color-color space and no indication of dust from visual inspection, similar to the blue UV-bright clusters in Figure \ref{FIG:reddening}.
There are 247 clusters (21.8$\%$) that are allowed a slightly higher level of reddening with a maximum E(B-V) = 0.5~mag. 
Clusters in these areas of the galaxy show small deviations from the models in the direction of reddening and a visual inspection shows low amounts of dust (green clusters in Fig. \ref{FIG:reddening}).
There are 301 clusters (26.6$\%$) which have a maximum adopted E(B-V) value of 1.0~mag. 
There is a noticeable amount of dust in these areas of the galaxy, leading to clusters falling to the right of the models (orange clusters in Fig. \ref{FIG:reddening}). 
Finally, one region comprised of 59 clusters (5.2$\%$) is given the highest maximum value of E(B-V) = 2.0~mag during SED fitting. 
This region is in the center of the galaxy and has a noticeably higher amount of dust than the other regions, seen in magenta in Figure \ref{FIG:reddening}.

\subsection{Age-Dating: Method, Results, and Checks}
\label{sec:agedating}

We estimate the age, reddening, and mass of clusters in NGC\,1614 by comparing the measured luminosities over 6 HST bands (NUV, U, B, V, H$\alpha$, and I) to predictions from the solar metallicity BC03 model with the maximum E(B-V) values described in $\S$\ref{sec:reddenings}. The H band and Pa$\beta$ filters are not used in the fit due to their poorer resolution and photometry.

NGC\,1614 has a value of 12 + log (O/H) = 8.69, consistent with solar metallicity \citep{Engelbracht2008,Modjaz2011}. 
Although any globular clusters will likely have sub-solar metallicity, we expect to detect very few of these ancient clusters at a distance of $\sim$70\,Mpc.
We also restrict our analysis in Section~4 to clusters younger than 0.5~Gyr, which are sufficiently young that we expect them to have approximately solar metallicity.

We use the BC03 models for age-dating.  While there are newer models which incorporate improved prescriptions for mass-loss \citep[e.g.,][]{Maraston05} and other models which include binaries and binary evolution \cite[e.g.,][]{Zackrisson11,Eldridge09}, we find that the BC03 models predict cluster colors which provide a better overall match to the observed colors of clusters.  For example, the PHANGS-HST collaboration found that the colors of $\sim100,000$ star clusters and associations in nearby spiral galaxies are well fit by the BC03 models.  In particular, \citet{Maschmann24} show that the measured (V-I) colors of clusters do not extend redward of the BC03 model predictions at 10~Myr.  However, a number of newer population synthesis models predict redder (V-I) colors than are observed \citep[e.g.,][]{Maraston05,Zackrisson11}.

In addition to photometry in broad-band filters, we directly include photometry measured in the F665N narrow-band filter (which is not continuum subtracted) in our fits. To predict the strength of H$\alpha$, we use the number of ionizing photons predicted by the BC03 models and assume Case B recombination to calculate the recombination line flux.  The predicted line emission is added to the stellar continuum at a given age in order to model the narrow-band filter.
Our default age-dating assumes that no ionizing photons escape ($f_{\rm esc}=0.0$), but we find that there is little impact on the age estimates if we assume $f_{\rm esc}=0.5$ instead.

H$\alpha$ emission is very strong for the youngest clusters and falls off quickly as they age.  The exact timescale for H$\alpha$ emission to essentially disappear is model-dependent.
The BC03 models predict this line emission is essentially gone by 6~Myr.  This is similar to predictions from Starburst99 \citep{Leitherer99}.
Models that include binaries like BPASS \citep{Eldridge09} can extend the life of H$\alpha$ emission to $\sim$10\,Myr when Wolf-Rayet stars are included, but this emission still peaks at ages $\lea$ 4\,Myr without them \citep{Dorn-Wallenstein18}. 
 
\begin{figure}
    \centering
    \includegraphics[width=0.40\textwidth]{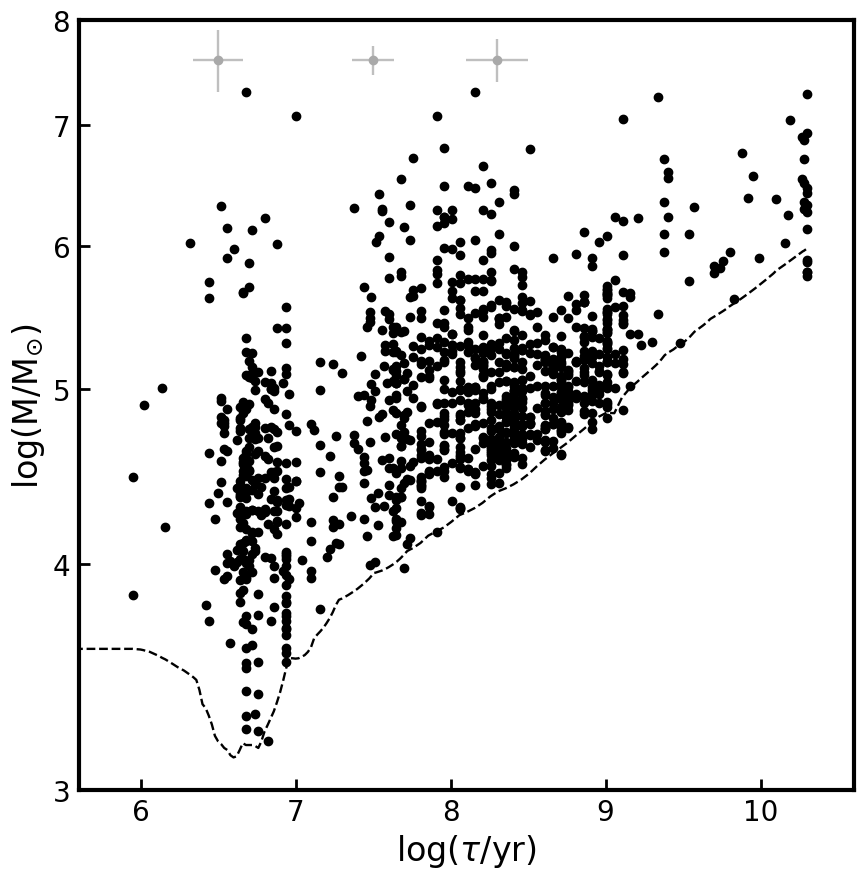}
    \caption{Cluster age–mass diagram for NGC\,1614. The dashed line represents M$_{\mathrm{V}}$ = -8 mag, the approximate completeness limit for individual clusters in our sample.
    Median error bars are plotted for clusters above the completeness limits for the 1-10\,Myr, 10-100\,Myr, and 100-400\,Myr age intervals (see text). 
    }
    \label{FIG:agemass}
\end{figure}

The grid for the SED fit runs over ages of log($\tau$/yr)=6.0 to 10.2 and reddening E(B-V)=0 to the allowed regional maximum (0.1, 0.5, 1.0, or 2.0~mag).
The best fit values for age and reddening for each cluster are found through minimizing the statistic: $\chi^2(\tau, A_V ) = \sum_{\lambda}~W_{\lambda} (m^{\mbox{obs}}_{\lambda}- m^{\mbox{mod}}_{\lambda})^2$ where $m^{\mbox{obs}}$ and $m^{\mbox{mod}}$ are the observed and model magnitudes respectively.
Masses for each cluster are estimated from the extinction-corrected V-band luminosity, and the age-dependent mass-to-light ratios predicted by the models, and the assumption that the distance modulus is 34.2~mag. 
We compute the $1\sigma$ error for each cluster
from the $\chi^2$ statistic.  We show the median uncertainty for clusters in the 1-10~Myr, 10-100~Myr, and 100-400~Myr age intervals at the top of Figure~\ref{FIG:agemass}.  These range from $\sim0.15$ to 0.2 in log~$\tau$, with the youngest and oldest age intervals having median $1\sigma$ errors around $\approx 0.2$, and the middle interval having somewhat smaller median errors $\approx 0.15$. 

Stochasticity is not expected to have much impact on cluster colors and hence on age and mass estimates, since the majority of 
clusters in NGC~1614 are quite massive with M $>$ $10^5$\,M$_\odot$. Stochastic fluctuations start to have a small affect on clusters with masses $\sim$3x$10^4$\,M$_\odot$ \citep{Goudfrooij21}, and become more pronounced at masses below $\sim$5x$10^3$\,M$_\odot$ in the blue and optical regimes we are probing here \citep{Fouesneau12}.

\begin{figure*}
    \centering
    \includegraphics[width=0.70\textwidth]{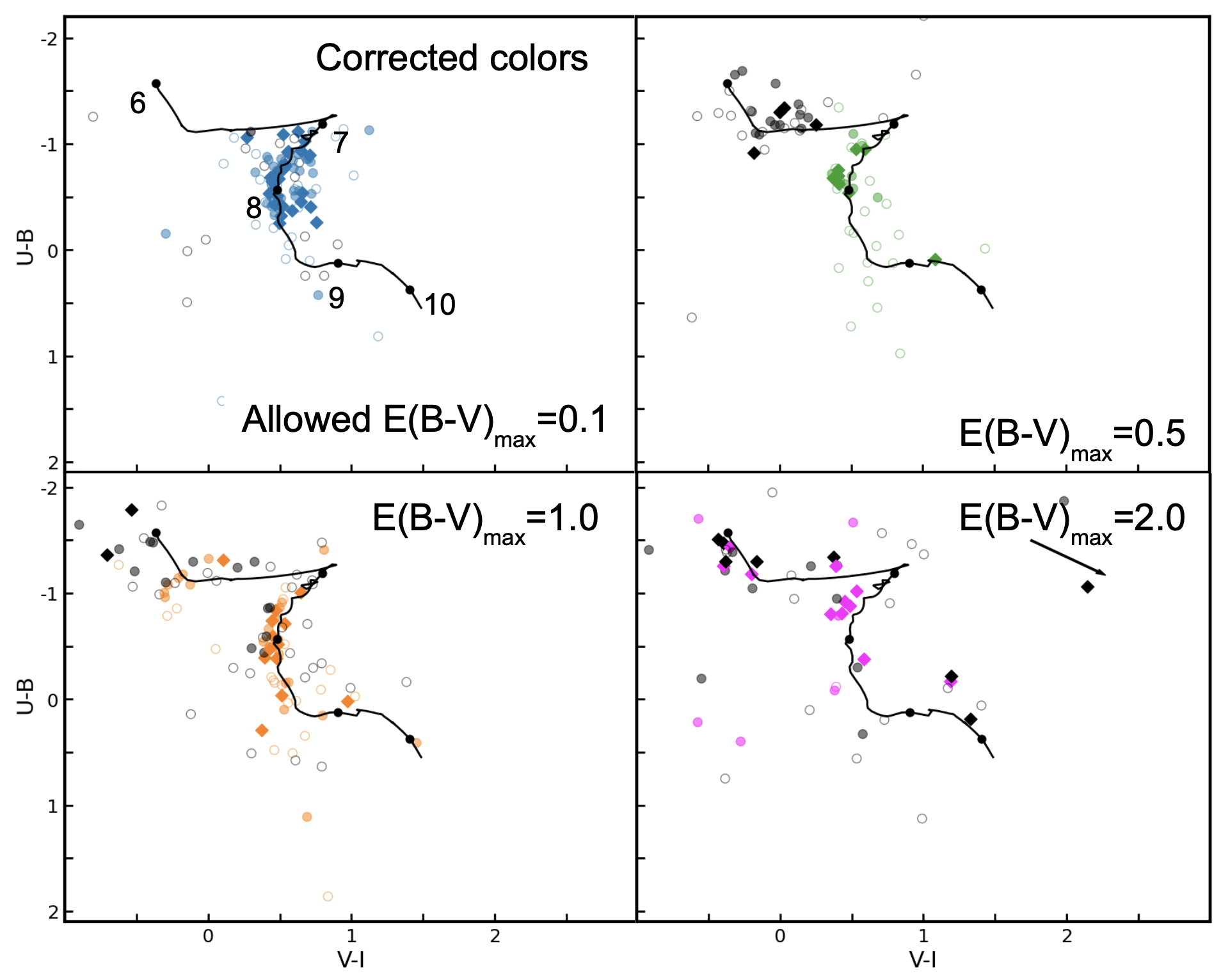}
    \caption{Same as the left of Figure \ref{FIG:reddening}, but with cluster colors corrected for best fit values from age-dating.
    }
    \label{FIG:unreddening}
\end{figure*}

Figure \ref{FIG:unreddening} shows U-B vs V-I color-color diagrams for the same four regions as Figure \ref{FIG:reddening}, but with cluster colors corrected by their best fit E(B-V) values after age-dating. For all regions, the H$\alpha$ bright clusters, shown in black, have moved to the top left of the models, where young clusters are expected to be. In the regions with larger E(B-V)$_\mathrm{max}$ values, clusters without H$\alpha$ stay along the older portion of the model track and do not move to young ages.

Ages of the bright clusters in the UV-bright arm, shown in blue in Figure \ref{FIG:unreddening}, have estimated ages between $\sim$25-250 Myr. However, if we allow the maximum E(B-V) to increase to 0.5 during age-dating for clusters in this region, the majority are best fit by too-young of an age ($\sim$6-10 Myr) and too-high reddening. 
The direct inclusion of H$\alpha$ in the fits prevents these (and other) clusters with no H$\alpha$ emission from being age-dated to less than $\sim$6 Myr, where the models predict the H$\alpha$ luminosity plummets. However, {\em it does not prevent them from having estimated ages between 7 and 10 Myr, which is why limiting the maximum E(B-V) is important.}

We compare our final age estimates with the results when assuming a single value of E(B-V)$_\mathrm{max}=1.5$~mag for the entire cluster population during age-dating, since this has been the default assumption for many studies \citep[e.g.,][]{Calzetti15,Adamo20,Lee22}
Not surprisingly, many clusters which were initially best fit by ages older than log($\tau$/yr) $>$ 7.5 using our variable maximum E(B-V) method have estimated ages younger than log($\tau$/yr) $<$ 7.0 when a higher maximum E(B-V) is allowed.  The vast majority ($\sim90$\%) of these clusters are fainter than $m_V=24$~mag, and none have associated H$\alpha$ emission.
The clusters which change age estimates are found throughout the galaxy, but are concentrated in areas with little on-going star formation and dust, like the UV-bright arm and tidal tails.  Therefore, we find that adopting a single value of E(B-V)$_\mathrm{max}$ = 1.5~mag results in incorrect (too young) age estimates for many clusters.
This issue has been pointed out in other recent studies as well.
Approximately 80\% of old globular clusters in the PHANGS-HST survey of nearby spiral galaxies were best-fit by ages that were too young by factors of 10-1000 \citep{Whitmore23b, Floyd24}.  A number of intermediate-age clusters were also best fit to ages $<10$~Myr (Thilker et al., submitted).

%---
\section{Results for Clusters}

\subsection{Age-Mass Diagram}

In Figure \ref{FIG:agemass}, we present the final age-mass diagram for clusters in NGC\,1614. Clusters range in age from log($\tau$/yr)$\approx$6 to 10.3 ($\sim$1~Myr to $>10$~Gyr) and have estimated masses between $\sim10^4$ and 2$\times 10^7~ M_{\odot}$. The magnitude limit of our observations is M$_{\mathrm{V}} = -8$~mag and shown as the dashed black line along the bottom edge of the age-mass diagram where we have assumed a distance modulus of 34.2~mag.
Above the luminosity limit of the sample, there are more lower mass clusters (M $\lea10^5$\,M$_\odot$) than high mass ones at any given age.
The luminosity limit also restricts us to higher mass clusters at older ages, because they fade over time.

The gap observed in cluster ages between log($\tau$/yr)$\sim$7.0 and 7.5 is due to well-understood biases from the age-dating process where the models loop back on themselves, and does not indicate an actual gap in cluster formation \citep{Gieles05,Goddard10,Chandar10b}.
There is a distinct lack of clusters younger than $\lea$3 Myrs. 
This is likely a bias in our age-dating procedure where it is hard to differentiate the ages of clusters between 1 and 4~Myr, often due to the amount of reddening that affects their broadband colors.
However, this bias should have little impact on the results of this paper. We bin all clusters from 1-10 Myrs into a single interval for our analysis of the CMF and $\Gamma$. Smaller bins are used in the age distribution, but the gaps are factored into the bins used there as well. 
We do not use clusters older than log($\tau$/yr)= 8.6 ($\sim 400$ Myr) in our analysis since at least some of these can be affected by the age-metallicity degeneracy \citep[e.g.,][]{Chandar04,Forbes22,Whitmore23b}.
We restrict the rest of our analysis to clusters younger than log log($\tau$/yr) $=8.6$ and brighter than $M_V=-8$~mag.

\subsection{Mass Function}
\label{sec:massfunc}

\begin{figure*}
    \centering
    \includegraphics[width=0.90\textwidth]{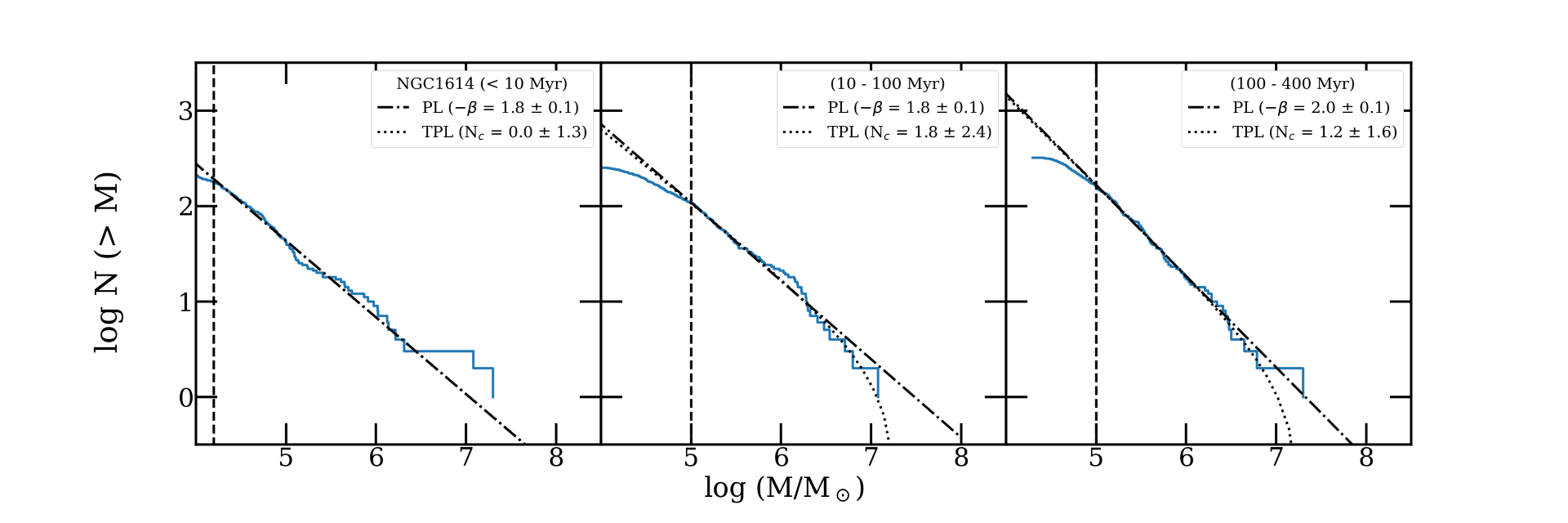}
    \caption{Cumulative mass functions for clusters in NGC\,1614 in age intervals of 1-10 Myr, 10-100 Myr, and 100-400 Myr. 
    The adopted completeness limit, where each distribution flattens from a power-law, is shown as the dashed vertical line.  Fits to a power-law fit a are shown as the dot-dashed lines and to a truncated power-law as the dotted line. The best fit values of $\beta$ for a power-law (PL) and the statistic $N_C$ which represents the upper end of a truncated power-law (TPL)} are given in the top right corner of each panel. See text for details.
    \label{FIG:MassFunction}
\end{figure*}

\begin{figure*} 
    \centering
    \includegraphics[width=0.90\textwidth]{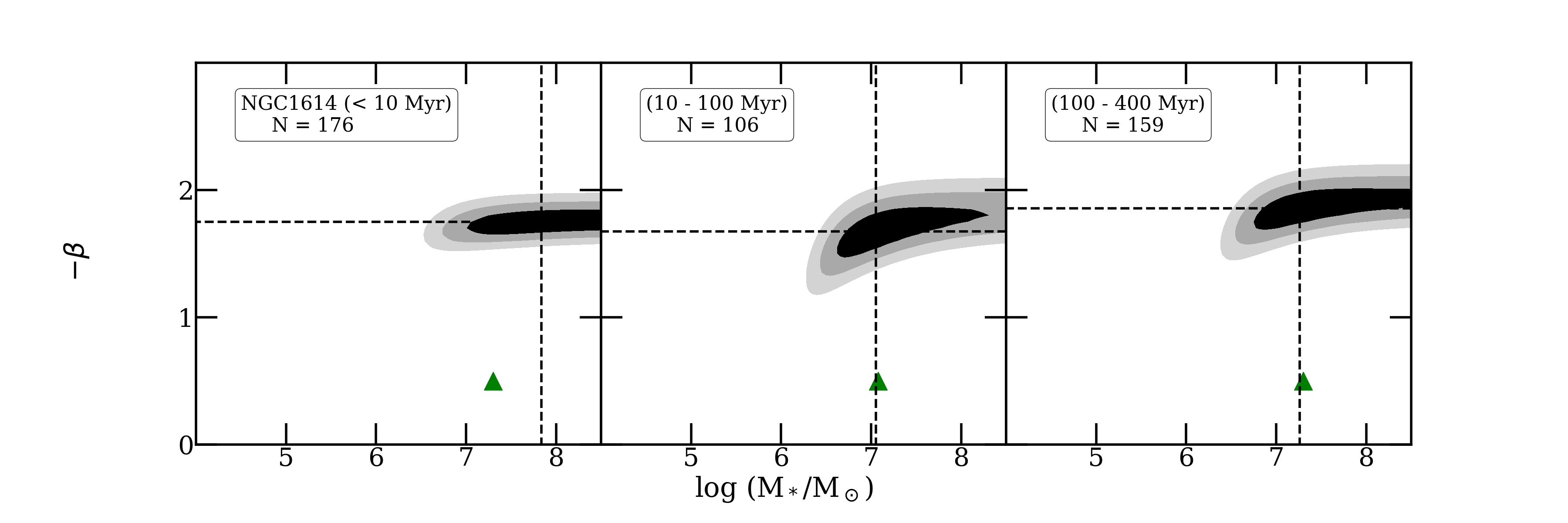}
    \caption{Maximum likelihood fit results to the power-law index $\beta$ and upper mass cutoff $M_*$ are shown for $<10$~Myr, $10-100$~Myr and $100-400$~Myr clusters. 
    The most massive cluster in each age range is plotted as the green triangle. The most likely -$\beta$ and M$_{\star}$ values are shown as the horizontal and vertical dashed lines respectively. 1$\sigma$ (black), 2$\sigma$ (grey), and 3$\sigma$ (light grey) confidence intervals are also shown.
    There is no convergence for M$_{\star}$ in any age range, indicating that the data do not prefer an exponential upper-mass cutoff. }
    \label{FIG:Likelihoods}
\end{figure*}

The shape of the cluster mass function (CMF) --- and how it changes over time --- can give critical insight into the formation and dissolution of clusters in a galaxy, for example, if there is a physical upper limit to the masses of clusters or whether mass-dependent or mass-independent disruption dominates cluster demographics.
The mass function of very young clusters with ages $<10$~Myr approximately represents the 'initial' cluster mass function (ICMF).  A comparison of the shape and normalization of this distribution with that of older clusters is important for understanding their evolution.

We study the cluster mass function in 3 different age intervals, 1-10 Myr, 10-100 Myr, and 100-400 Myr.
Figure \ref{FIG:MassFunction} shows the cumulative cluster mass function in each of the three age intervals.  
For each one, the distribution increases in an approximately power-law fashion before flattening towards the lower mass end. 
The flattening at the lower mass end of the CMF is assumed to be due to sample incompleteness, and not to a physical effect.
We determine the completeness limit for each distribution as the mass where the distribution deviates below a power-law at the 99$\%$ level (dashed vertical line).
The completeness limits are found to be log(M/$\ M_{\odot}$) = 4.2 ($<$ 10~Myr), 5.0 (10-100 Myr), and 5.0 (100-400 Myr).

The best fit single power law with dN$/$dM $\propto M^{\beta}$ down to the completeness limit is shown as the dot-dashed line in each panel.
These fits give $\beta$ = -1.8, -1.8, and -2.0 $\pm$ 0.1 for 1-10 Myr, 10-100 Myr, and 100-400 Myr respectively. 
This means that {\em the CMFs can be described by a single power-law with index $\beta=-1.9\pm0.1$ for NGC\,1614, and there is no obvious evolution in the shape of the mass function over the first $\approx 0.5$~Gyr}.

Some previous works have found that cluster mass functions are significantly better described by a Schechter function than by a single power law \citep{Bastian12,Johnson17}.
A Schechter function has the form: $\psi(\mathrm{M}) = d\mathrm{N}/d\mathrm{M} \propto \mathrm{M}^{\beta} \exp(-\mathrm{M}/\mathrm{M}_{\star})$, a power-law with an index $\beta$ with an exponential cutoff at a value $\mathrm{M}_{\star}$ at the upper end.  
To test for a potential upper cutoff in the mass function, we use the maximum-likelihood method developed and described in \cite{Mok19}. 
This method uses all clusters above the completeness limit, and does not bin the clusters or use a cumulative distribution to smooth over any features. 
The best fit for these parameters, shown by the dashed lines, is obtained by maximizing the likelihood function. The 1, 2, and 3$\sigma$ confidence contours can be defined in the $\beta$-M$_*$ plane using the formula: $ln(\mathrm{L}) = ln(\mathrm{L}_{\mathrm{max}}) - 1/2 \chi^2_\mathrm{p} (\mathrm{k})$, where p is the confidence level and k is equal to the number of free parameters in the fit.

The results for our maximum-likelihood fits to a Schecter function are plotted in Figure \ref{FIG:Likelihoods} for each age range. 
These plots show the 1$\sigma$, 2$\sigma$, and 3$\sigma$  confidence intervals (black, grey, and light grey, respectively) for the maximum-likelihood fit of the power law index -$\beta$ and the cutoff mass log($\ M_{\star}$/$\ M_{\odot})$. 
The most massive cluster observed in each age interval is plotted as a green triangle. 
A wide range of $\ M_{\star}$ values is allowed by each distribution, because the $2\sigma$ and $3\sigma$ confidence contours start near 10$^{6.5}$ $\mathrm{M}_{\odot}$ and continue without closing to the right edge of the diagrams at 10$^{8.5}$ $\mathrm{M}_{\odot}$ for all three intervals, and even the $1\sigma$ contour is open all the way to the edge for $1-10$~Myr clusters.
This means that {\em the CMFs are consistent with being drawn from a pure power law rather than requiring a Schecter-like cutoff at the upper end}.

We note that uncertainties for the mass estimates were not included in the maximum likelihood fits.  Including uncertainties in the fits would further broaden the contours, making any detection of M$_{\star}$ even less significant.

As an independent check on the results from the maximum-likelihood method, we also fit a broken power-law to the cumulative mass distributions in Figure~\ref{FIG:MassFunction}.  
A broken power-law allows a more gradual downturn than an exponential and acts like a simplified Schechter function.
This is described by:
N (M$^{\prime}$ $>$ M) = N$_c$ [(M/M$_{\star}$ )$^{\beta +1}$ - 1]
where N$_c$ must be statistically significant ($\ge$ 3$\sigma$) to indicate a truncation better represents the upper end of the distribution than a single power law.  The best truncated power-law fits to the CMFs are shown as the dotted lines in Figure \ref{FIG:MassFunction}. 
None of the N$_c$ values reach 3$\sigma$, indicating that a truncated power-law is also not a good fit to the cluster mass functions in NGC\,1614 in any of the age intervals studied here.

At the distance of $\sim$70\,Mpc, it is possible that multiple clusters could blend together and appear as a single, massive cluster instead. 
\cite{Randriamanakoto13} experimented on the nearby ($\sim$20\,Mpc) 
merging Antennae galaxies to test the effects of blending on super star clusters. They degraded HST images to mimic how the Antennae would look at a distance of 70\,Mpc, nearly the exact same distance as NGC~1614. They found that while blending can flatten the observed cluster luminosity function, the effect is minor and barely detectable outside of the fit uncertainties.
We have run a few experiments along the lines of those performed in \cite{Randriamanakoto13} and
\cite{Chandar23b} to see if blending has a strong impact on our results. 
These experiments include: (1) splitting the most massive cluster in each age bin into 3 equal-mass ones, and (2) taking 10 random clusters from the 50 most massive and splitting them into 3 equal-mass clusters, or (3) deleting them from the catalog. The mass functions for all of these experiments have best fit values of the power-law index $\beta$ that are within the errors of our initial fit and none show any indictation of a Schechter-like cutoff.

For NGC\,1614, we find no significant detection of an upper mass cutoff in the CMFs, and that the shape of the CMF does not evolve over the first $\approx0.5$~Gyr over the plotted range of masses.

\subsection{Age Distribution}
\begin{figure}
    \centering
    \includegraphics[width=0.40\textwidth]{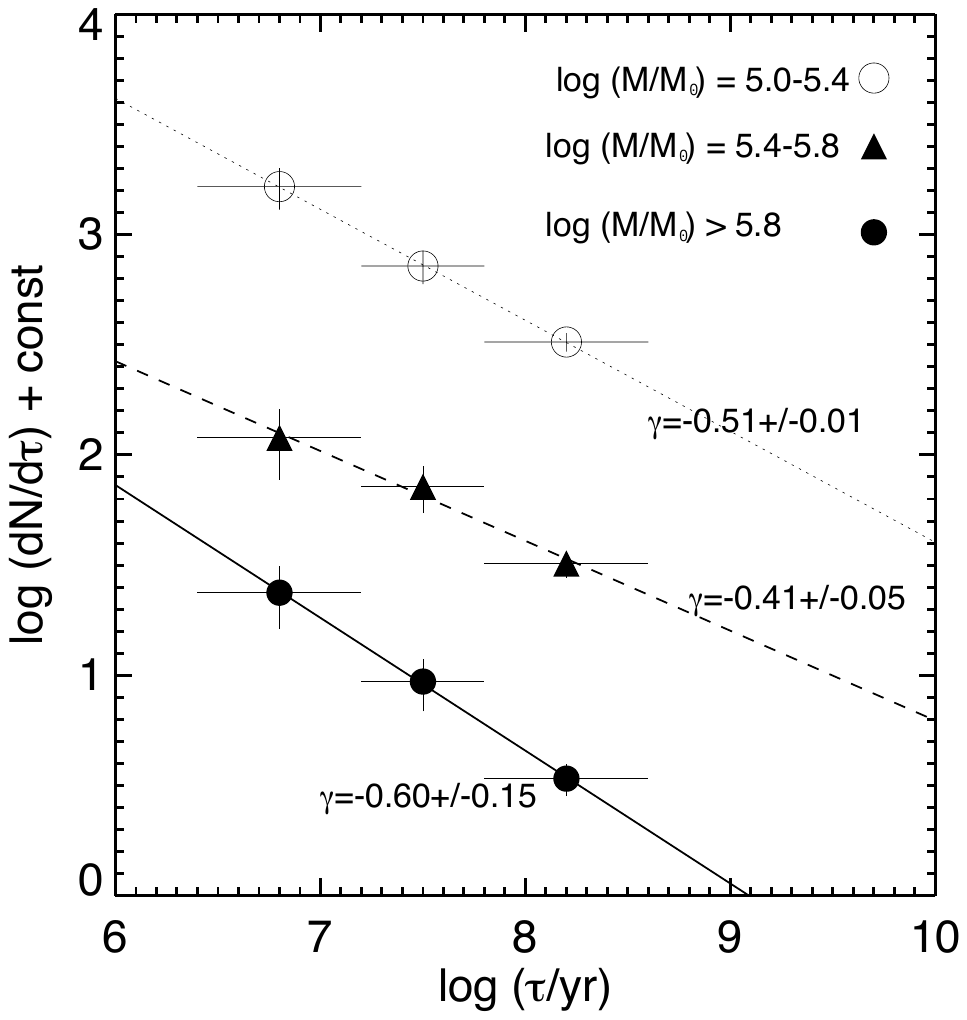}
    \caption{Age distributions of star clusters in NGC\,1614 in the 3 indicated mass intervals. Fitted lines show power laws, dN/d$\tau$ $\propto$ $\tau^\gamma$, with the best fit value of $\gamma$ indicated. }
    \label{FIG:agedist}
\end{figure}

Cluster age distributions encode important information on the formation and disruption histories of the clusters.
Cluster age distributions for NGC\,1614 are plotted in Figure \ref{FIG:agedist} for three mass intervals: log(M/M$\ _{\odot}$) $\ge$ 5.8, log(M/$\mathrm{M}_{\odot}$) = 5.8-5.4, and log(M/M$\ _{\odot}$) = 5.0-5.4.  
All of these clusters have luminosities that are above the completeness limit.
These are plotted as the number of clusters within the age bin vs the amount of time covered by the bin.
We have used fairly broad bins in log($\tau$) to smooth out the small-scale features and age-dating artifacts seen in the age-mass diagram (Figure \ref{FIG:agemass}). 
.

We model each age distribution as a power-law, $dN/d\tau \propto \tau^{\gamma}$, and fit for the power-law index $\gamma$. Horizontal bars are the width of the age bin, and vertical error bars are calculated from Poisson statistics.
We find that the age distributions for all three mass intervals decline continuously starting at young ages and are similar within the uncertainties.  This means that the age distributions are independent of their masses, at least in the mass ranges probed here.  We find that a power-law index of  $\gamma \approx -0.5\pm 0.1$ describes the cluster age distributions in NGC\,1614.

We expect that there are at least some very young, recently formed clusters that remain obscured at optical wavelengths but that can be detected by high-resolution imaging in the infrared with JWST, similar to the clusters identified in $\S$ \ref{sec:missingIR}. This population of young, embedded clusters might further steepen the power-law index for the cluster age distribution if they were included, an effect that was observed recently by \cite{Linden23} for the dusty, merging system VV\,114.
This very young, optically obscured cluster population will be explored using upcoming JWST observations.

\subsection{Fraction of Stars in Clusters}
\label{sec:gammamethod}

The fraction of stars that are born in clusters, also known as the cluster formation efficiency ($\Gamma$), is a fundamental property of star and cluster formation on galaxy scales.  
Simulations mostly find that $\Gamma$ increases significantly with $\Sigma_{\rm SFR}$.
Typically, $\Gamma$ is calculated from the stellar mass found in very young $1-10$~Myr clusters, which we refer to as $\Gamma_{\rm 1-10~Myr}$.  Fewer works have estimated the fraction of stars that remain in clusters at older ages.
In this section, we calculate the fraction of stars found in clusters within NGC\,1614 as an additional data point at the high end of SFR and $\Sigma_{\rm SFR}$.  We track this fraction from very young, $1-10$~Myr clusters, to those in older $10-100$~Myr and $100-400$~Myr clusters.

In order to calculate the fraction of stars in clusters in any age interval, we must calculate both the total mass of stars born during that time interval and that found within the compact cluster population. We summarize our method below, which follows standard practices developed initially in \citet{Goddard10} and used in a number of recent works \citep[e.g.,][]{Adamo15,Chandar17,Adamo20,Cook23}.

To find the total mass of stars in clusters, we first sum up the masses of all observed clusters above the completeness limit of our sample (see Sec. \ref{sec:massfunc} and Fig. \ref{FIG:MassFunction}). 
The total stellar mass in clusters below the completeness limit is determined by extrapolating the CMF from the completeness limit down to 100 $\mathrm{M}_{\odot}$ assuming a power-law with an index of $-2.0$.  This power-law index is similar to the best fits we determined for the cluster mass functions in NGC\,1614 (see $\S$ \ref{sec:massfunc}).
The power law is integrated and added to the mass in clusters above the completeness limit to get the total mass of stars in clusters. 

The total mass of stars in the galaxy is found by multiplying the total SFR of the galaxy by the time elapsed in the age interval.  $\Gamma$ is then simply the ratio of the mass of stars in clusters divided by the total stellar mass.
For our calculations, we adopt a SFR of 49.6 $\ M_{\odot} yr^{-1}$ \citep{Gimenez-Arteaga22} for NGC\,1614 (see $\S$~\ref{SEC:sfr}) with an uncertainty of $\sim25$\% (following \cite{Cook23} and \cite{Chandar23b}).

We calculate the fraction of stars that are born in clusters to be $\Gamma_{\rm 1-10~Myr} = 22.4 \pm 5.7$\%.
We find the fraction of stars that remain in the 10-100 and 100-400 Myr age intervals to be: $\Gamma_{10-100}$ = 4.5$\%$ $\pm$ 1.1$\%$, and $\Gamma_{100-400}$ = 1.7$\%$ $\pm$ 0.4$\%$.

\section{Discussion}

\subsection{No Cut-Off in the Cluster Mass Function}

The observed CMF can be used to infer maximum cluster masses and other properties such as $\Gamma$. 
It has been suggested that the CMFs of galaxies have a similar power-law index of $\beta$ = -2 (e.g., \cite{Zhang99,fall12,Chandar17,Krumholz19,Li17,Li18,Grudic21}), but with high-mass cutoffs (M$_*$) that increase with $\Sigma_{SFR}$ \citep{Johnson17}. 
Recent simulations both recreate and utilize the the physics driving the CMF. Tests constraining the ICMF include magneto hydrodynamical simulations of turbulent, star forming giant molecular clouds (GMCs) \citep{Grudic21,Grudic23}, while cosmological simulations have included star clusters as a unit of star formation in high redshift Milky Way-sized galaxies \citep{Li17,Li18}.

With the adopted SFR for NGC\,1614 from \cite{Gimenez-Arteaga22} of 49.6~M$_{\odot}$~yr$^{-1}$, we can use the relationships from the fiducial run in \cite{Li17} of M$_*$ $\approx$ 1.4 x 10$^4$ M$_\odot$ x SFR$^{1.6}$ and M$_\mathrm{max}$ $\approx$ 8.8 x 10$^4$ M$_\odot$ x SFR$^{1.4}$ to calculate the most massive cluster (M$_{\mathrm{max}}$) and M$_*$ predicted in NGC\,1614. 
M$_*$, according to the relation, is calculated to be 7.2 $\times$10$^6$~M$_{\odot}$ and 
M$_{\mathrm{max}}$ to be 2.1 $\times$ 10$^7$~M$_{\odot}$ for a galaxy the SFR NGC\,1614.

The most massive cluster observed in NGC\,1614 is 2.0 $\times$ 10$^7$~M$_{\odot}$, which agrees well with the predicted value from \cite{Li17}.
However, as presented in $\S$ \ref{sec:massfunc}, no high-mass cutoff is found in the CMF in this galaxy. 
Therefore, our results do not agree with the predicted cutoff value of 7.2 $\times$ 10$^6$~M$_{\odot}$ from \cite{Li17}. 

As our results only probe $\sim$ 3$\%$ of the Hubble time, it is reasonable to assume that the physics of cluster formation has not changed significantly over the past 400 Myr.

\subsection{Constraints on Cluster Formation and Disruption}

The study of cluster formation is entwined with cluster disruption. 
Based on global estimates of the star formation rate from tracers which are sensitive to different age intervals, we concluded that the star formation history of NGC\,1614 has likely been fairly constant (within a factor of $\approx2-3$) for the past $\sim0.5$ Gyr (see $\S$ \ref{SEC:sfr}).
A fairly constant star formation history is also supported by the continuous distribution of cluster colors along the cluster evolutionary track
(Figure \ref{FIG:CC}), which does not show gaps or concentrations in any particular age interval.
These distributions contrast with the color distributions and age-mass diagrams for a post-starburst galaxy like S12 \citep{Chandar21}, and NGC~34, which is a LIRG that is post-merger and appears to be post-burst (Zhang et al., in prep).  All of these points support our assumption that NGC\,1614 has been forming stars and clusters at a fairly constant rate.

\begin{figure*}
    \centering
    \includegraphics[width=0.95\textwidth]{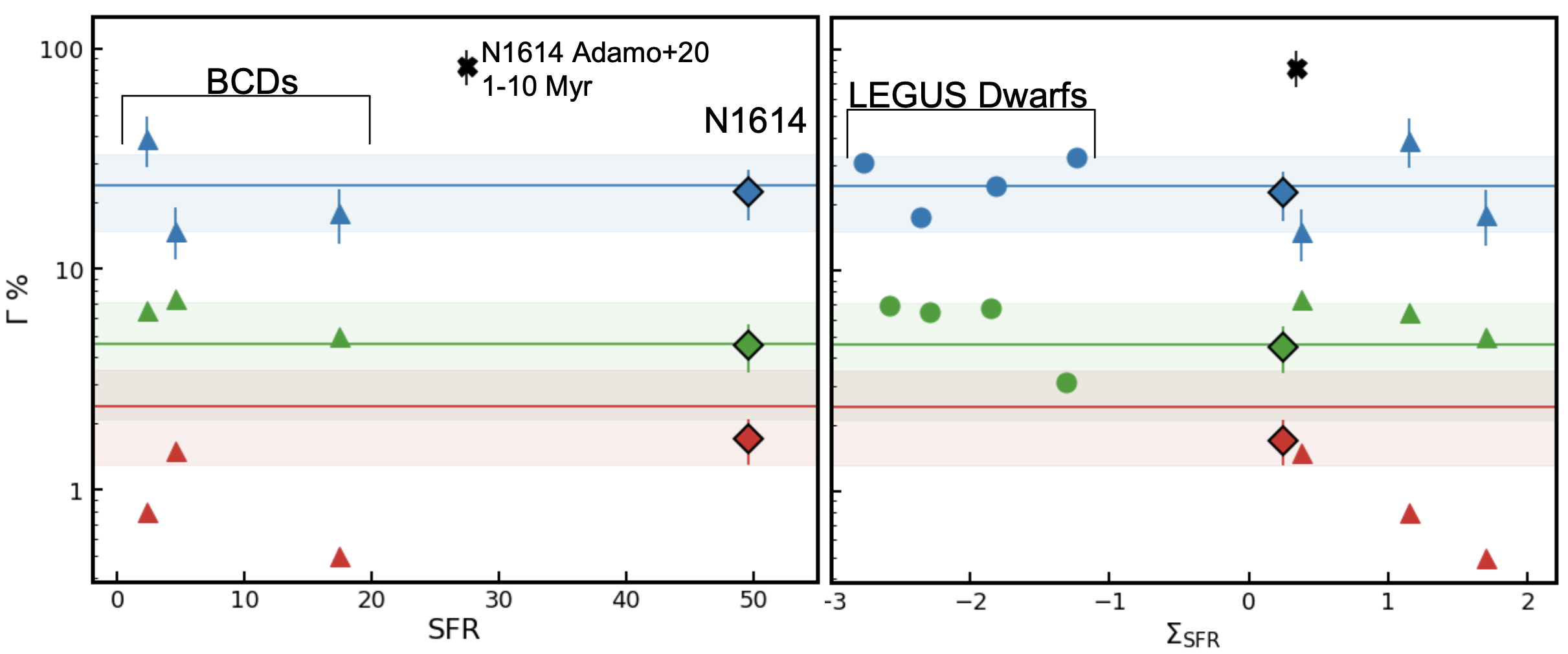}
    \caption{Comparison of $\Gamma$ values of different galaxies and age intervals. $\Gamma_{1-10}$ is plotted in blue, $\Gamma_{10-100}$ in green, and $\Gamma_{100-400}$ in red. Values for NGC\,1614 in this work are plotted as diamonds, $\Gamma_{1-10}$ for NGC\,1614 from \cite{Adamo20} are plotted as black Xs. Blue compact dwarf galaxies from \cite{Chandar23b} are plotted as triangles. Binned dwarfs in the LEGUS survey from \cite{Cook23} are plotted as circles. The relationships found in \cite{Chandar17} are the lines with shaded error regions in each color. }
    \label{FIG:gammaHi}
\end{figure*}

\subsubsection{Cluster Formation}
$\Gamma_{1-10}$ is a tracer of the cluster formation efficiency, and measures the percentage of stars that are born in clusters.  In NGC\, 1614, an extreme system with a high SFR, $\Gamma_{1-10}$ can tell us about galactic scale star formation and if cluster formation is more efficient in more extreme systems. 

Simulations of $\Gamma_{1-10}$ over the past decade vary in approach and assumptions, but all predict an increase in $\Gamma_{1-10}$ with $\Sigma_{\rm SFR}$.
These simulations range from analytical scaling predictions \citep{Kruijssen12}, to magneto hydro-dynamical simulations of GMCs and model cluster populations.
Different prescriptions for feedback can dramatically impact the $\Gamma -\Sigma_{\rm SFR}$ relation \citep[e.g.,][]{Li18,Grudic22,Dinnbier22}.
The predicted increase over a factor of $\sim10^4$ in $\Sigma_{\rm SFR}$, ranges from a factor of $\sim$100x \citep{Kruijssen12} to only $\sim$ 2 \citep{Dinnbier22}.
For a system like NGC\,1614, the simulations predict a $\Gamma_{1-10}$ between 3 and 60~$\%$, with most predictions in the 20-50$\%$ range. 
This large range in predicted values underscores the importance of empirical estimates of $\Gamma_{1-10}$ in galaxies with high $\Sigma_{\rm SFR}$.

We calculate the fraction of stars that form in clusters to be $\Gamma_{\rm 1-10~Myr} = 22.6 \pm 5.7$\%. 
This fraction may be somewhat higher if there is a significant number of massive, deeply embedded young clusters, but our NIR images ($\S$ \ref{sec:missingIR}) tentatively indicate optically-obscured clusters are likely to have lower masses and hence not to have a significant impact on $\Gamma_{1-10}$. 
Our $\Gamma_{\rm 1-10~Myr}$ of 22.6 $\pm 5.7\%$ agrees well with the range of $\Gamma_{1-10}$ = 24$\%$ $\pm$ 9$\%$ found in eight galaxies which range from dwarfs to spirals to mergers \citep{Chandar17}.
It is similar to the results found for 23 nearby dwarf and irregular galaxies studied as part of the LEGUS survey of $\Gamma_{1-10}$ = 27$\%$ $\pm$ 6$\%$ \citep{Cook23}.
This value is also similar to that found for three blue compact dwarf galaxies from the CCDG sample, dwarf galaxies with some of the highest $\Sigma_{\rm SFR}$ in the nearby universe \citep{Chandar23b}.

On the other hand, some previous works have found significantly higher values for $\Gamma_{1-10}$ for galaxies with high $\Sigma_{\rm SFR}$, including for NGC\,1614 \citep{Adamo20}. Our calculated $\Gamma_{1-10}$ in NGC\,1614 is nearly a factor of four times lower than the value of $\Gamma_{1-10}$ = 83.1$\%$ $\pm$ 15.2$\%$ found by \cite{Adamo20} for this galaxy; this discrepancy --- mostly caused by assumptions in SFR and details of the age-dating method --- is discussed further in the Appendix. Note that \cite{Adamo20} did not calculate $\Gamma_{10-100}$ or $\Gamma_{100-400}$.

\subsubsection{Cluster Disruption}
The fraction of stars that remain in older clusters provides important constraints on the disruption of the clusters.
As we believe there has been a fairly consistent star formation history in NGC\,1614, we can compare values of $\Gamma_{\rm 1-10~Myr}$ with $\Gamma_{\rm 10-100}$ and $\Gamma_{\rm 100-400}$ to probe the dissolution of clusters within NGC\,1614.
We find: $\Gamma_{\rm10-100}$ = 4.2$\%$ $\pm$ 1.1$\%$ and $\Gamma_{\rm100-400}$ = 1.4$\%$ $\pm$ 0.4$\%$.
These values show the mass of stars in clusters decreases significantly, indicating that clusters begin to disrupt starting soon after they are born, and continues for at least the first $\sim0.5$~Gyr.

The shape of the cluster age distribution in NGC\,1614 supports this picture of early, continuous cluster disruption. 
As shown in Figure \ref{FIG:agedist}, 
the age distributions of clusters in NGC\,1614 are fairly similar in shape across the different mass intervals while staying above the completeness limit. 
The declining shape starts soon after formation, with a power-law index of $\gamma$ $\approx$ -0.5 for all plotted masses.  A power-law index of $-0.5$ indicates that $\approx70$\% of clusters are disrupted each factor of ten in age. We do not see evidence of lower mass clusters being disrupted earlier than higher mass ones through the first $\sim$0.5 Gyr that we study here. 
Our estimates for $\Gamma$ and the shape of the cluster age distribution agrees with those found in previously published works for other star forming galaxies \citep{Chandar17,Cook23}.

Our $\Gamma_{\rm10-100}$ = 4.2$\%$ $\pm$ 1.1$\%$, and $\Gamma_{\rm100-400}$ = 1.4$\%$ $\pm$ 0.4$\%$ values again agree well with the relation over a range in galaxies found by \cite{Chandar17} of $\Gamma_{10-100}$ = 4.6$\%$ $\pm$ 2.5$\%$, and $\Gamma_{100-400}$ = 2.4$\%$ $\pm$ 1.1$\%$.  
\cite{Cook23}, in a binned sample of more than 20 nearby dwarf galaxies, only calculated $\Gamma_{10-100}$ = 7$\%$ $\pm$ 2$\%$, which is also in agreement with our calculated value.
\cite{Chandar23b} found ranges of $\Gamma_{10-100}$ = 5 - 8$\%$ and  $\Gamma_{100-400}$ = 1-2$\%$ for their three blue compact dwarf galaxies.

\subsection{Constraints on Ages in the UV-Bright Arm}
\begin{figure*}
    \centering
    \includegraphics[width=0.95\textwidth]{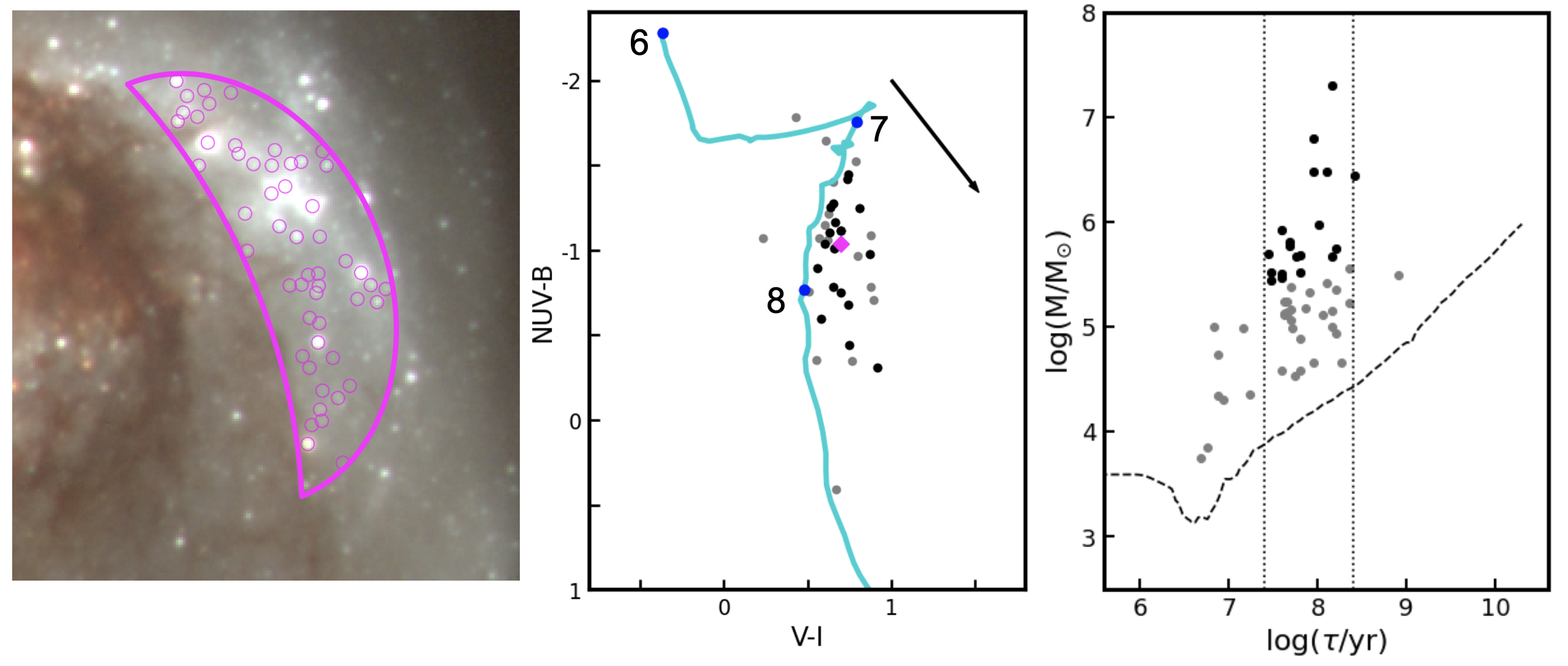}
    \caption{Clusters in the UV-bright arm of NGC\,1614. \textbf{Left:} BVI HST color image with clusters identified by the magenta circles. \textbf{Middle:} NUV-B vs V-I color-color diagram of clusters brighter than m$_\mathrm{V}$ $\le$ 23 shown in black. The median color of these clusters is shown as the magenta diamond. An A$_\mathrm{V}$ = 1.0 reddening vector is shown in the top right. \textbf{Right:} Age-mass diagram of clusters within the UV-bright arm. The vertical dashed lines are at 25 and 250 Myrs.}
    \label{FIG:UV_bright_plots}
\end{figure*}

Star-forming clumps, often referred to as "star forming complexes", have $\sim$kpc sizes and can be seen in galaxies out to redshifts of z$=4-5$, although substructure and individual clusters cannot be resolved at these distances \citep{Elmegreen05,Elmegreen09,Guo15,Guo18}.
High-redshift clumps tend to be very massive, $\sim$ 10$^8$-10$^9$M$_{\odot}$ which is $\sim$100x higher than in local, non-interacting galaxies \citep{Elmegreen09}.
However, local interacting galaxies, like the Antennae \citep{whitmore10}, show signs of higher star formation rates and larger cluster and clump sizes.

In the local universe, star-forming clumps have different properties in interacting vs. non-interacting galaxies.
In a study of more than 1000 clumps selected from 8$\micron$ Spitzer images across 46 interacting and 38 non-interacting spirals within 70~Mpc, \cite{Zaragoza18} found that clumps in interacting galaxies have higher $\Sigma_{\rm SFR}$ and younger ages than those in non-interacting galaxies, based on integrated photometry SED fitting of the clumps in the NUV through IR. 
From the CCDG HST sample of nearby interacting galaxies and blue compact dwarfs, 
\cite{Elmegreen21} found $\sim$50 clumps across the sample after degrading the observations to mimic galaxies at up to z = 2. SED-based estimates for the clumps at low redshift found an average age of $\sim$180~Myr - $\sim$500~Myr.

The UV-bright arm on the west side of the late-stage merger NCG\,1614 shows little-to-no H$\alpha$ emission and therefore has not been experiencing star formation for at least the past $\approx10$~Myr. 
The brightest region covers $\approx4$~kpc$^2$, outlined in Figure~\ref{FIG:UV_bright_plots}, and is representative of clumps found in galaxies at higher redshift \citep{Elmegreen21}.
The clusters identified in this region are circled in magenta and have colors that closely follow the BC03 solar metallicity track, indicating they experience little reddening.
The colors indicate that massive clusters formed mostly $\sim$250 to 50~Myr ago, shown by the dashed vertical lines in the right panel of Figure~\ref{FIG:UV_bright_plots}.
We find the median age of this region to be log($\tau$/yr) = 7.79 ($\sim$60~Myr) with a standard deviation of 0.28 for clusters brighter than m$_V$ = 23~mag within the contours shown in Figure \ref{FIG:UV_bright_plots}.
Median V-I and NUV-B colors are found to be 0.68~mag and -1.1~mag respectively, and are shown as the magenta diamond in Figure \ref{FIG:UV_bright_plots}. 
There is one very bright cluster in this region with an estimated mass around 10$^7$ M$_{\odot}$; the total mass of clusters with $M \geq 10^5$M$_{\odot}$ is $\sim5.0\times10^7~\mathrm{M}_{\odot}$.

The duration of star and cluster formation in the UV-bright arm in NGC\,1614 appears similar to that for hinge clumps, which are usually found in the tidal features of merging galaxies. 
\cite{Smith14} found sustained star formation in hinge clumps of 5 nearby interacting galaxies, occurring in either multiple bursts or for a significant duration, rather than in a single, short-lived burst.  This analysis was based on multi-wavelength observations from the far-UV through X-ray, including H$\alpha$ line strength.
The prolonged star formation in hinge clumps is likely due to a prolonged inflow of gas.
Massive clusters in the UV-bright arm of NGC\,1614 have an age spread of $\approx200$~Myr, indicating that star formation in this region was also prolonged rather than a single, short-lived burst.

\section{Summary and Conclusions}

In this work, we studied the star cluster population in the luminous infrared galaxy NGC\,1614 using HST photometry in 8 bands: NUV (F275W), U (F336W), B (F438W), V (F555W), H$\alpha$ (F665N), I (F814W), Paschen$\beta$ (F130N), and H (F160W). 
A key goal was to obtain accurate age, reddening, and mass estimates for the clusters, which requires successfully breaking the age-reddening degeneracy, as a means to allow for a cluster population analysis of the rest of the CCDG sample.
We used an updated method to break this degeneracy, utilizing a comparison of predicted and observed SEDs of clusters in the 6 optical bands, including the narrow-band H$\alpha$, in which
we scale the maximum E(B-V) allowed during the fitting procedure by the amount of dust in the region.

\begin{itemize}
    \item The distributions of cluster colors in NGC\,1614, as seen in Figures \ref{FIG:CC} and \ref{FIG:reddening}, are found to be fairly continuous and show a range of reddening in different locations, from very low in the UV-bright arm to an E(B-V)$\approx$2.0~mag in the dusty central region.
    \item The most massive clusters have M$\approx \mbox{few} \times 10^7~\mathrm{M}_{\odot}$, comparable to those found in other extreme systems, like the Antennae.
    \item No statistically significant high-mass cut-off in the cluster mass function was found. In addition, maximum likelihood fits of cluster masses for all studied age intervals (1-10~Myr, 10-100~Myr, 100-400~Myr) are found to be well fitted by a power law of $\sim$-1.8, and do not show statistically significant evidence for a Schechter-like upper mass cutoff.
    \item  The fraction of stellar mass born in clusters was calculated from the $1-10$~Myr clusters, and found to be $\Gamma_{1-10}$ = 22.4$\%$ $\pm$ 5.7$\%$.  This value is similar to values found for galaxies with 
    $\Sigma_{\rm SFR}$s $\sim$1000 times lower than NGC~1614.
    \item The fraction of stars that remain in clusters surviving to ages of 10-100~Myr and 100-400~Myr are found to be $\Gamma_{10-100} = 4.5\pm 1.1$\% and $\Gamma_{100-400}=1.7\pm 0.4$\%, respectively. These results indicate that cluster disruption begins soon after the clusters form and continues for at least the first $\sim$.5 Gyr.
    \item The early, rapid dissolution of clusters is supported by the age distribution, which can be described by a simple power-law with an index $\approx-0.5 \pm 0.1$ for clusters with masses greater than $10^5~M_{\odot}$ up to ages of at least $\approx0.5$~Gyr.
    \item The UV-bright arm has properties similar to stellar clumps observed in galaxies at redshift z$\approx2$, and experienced fairly constant star formation for a period of $\approx$200~Myr starting 250~Myr ago.
\end{itemize}

%--------------------------------
\begin{acknowledgements}
R.C. acknowledges support from HST-GO-15649, and we thank the anonymous referee for suggestions which improved our manuscript.
The HST data presented in this article were obtained from the Mikulski Archive for Space Telescopes (MAST) at the Space Telescope Science Institute. The specific observations analyzed can be accessed via \dataset[10.17909/3yqt-mh67]{https://doi.org/10.17909/3yqt-mh67}


%\bibliographystyle{aas}   

\end{acknowledgements}

\software{Photutils \citep{photutils}, 
    Matplotlib \citep{matplotlib}, 
    NumPy \citep{numpy-guide,numpy},
    Astropy \citep{astropy22},
    APLpy \citep{Robitaille12}
    SciPy \citep{scipy_new},  
    SAOImage DS9  \citep{ds9}
          }
\bibliography{master}
\bibliographystyle{aasjournal}

%%%%%%%%%%%%%%%%%%%%%%%%%%%%%%%%%%%%%%%%%%%%%%%%%%

\clearpage
%\onecolumn

%%%%%%%%%%%%%%%%% APPENDICES %%%%%%%%%%%%%%%%%%%%%

\appendix
\section{The Dependence of $\Gamma$ on Assumptions}

The fraction of stars born in clusters, $\Gamma_{1-10}$, gives key insight into how efficient the cluster formation process is in different star-forming environments. $\Gamma$ is defined as:
\begin{equation}
    \Gamma = \frac{\mathrm{Mass\;of\;Stars\;in\;Clusters}}{\mathrm{Mass\;of\;All\;Stars}}
\end{equation}
where the denominator, the total mass in stars, is simply calculated from the star formation rate (SFR) multiplied by the age interval. The numerator, the mass of stars found in clusters, is calculated in two parts from (1) the sum of all masses of observed clusters above the completeness limit (within the age bin; see $\S$\ref{sec:massfunc}) plus (2) the total mass calculated by integrating a power-law with index $\beta$=-2 over the mass range from 100 $\mathrm{M}_{\odot}$ up to the cluster completeness limit (see $\S$\ref{sec:gammamethod} for more details).

The calculation of $\Gamma$ relies on a number of assumptions which can drastically affect the result in some cases, and therefore the physical interpretation.
Here, we explore the different assumptions which have led to different estimates of $\Gamma_{1-10}$ for NGC~1614.

One key assumption that can strongly affect $\Gamma$ is the assumed SFR. 
Estimates of the SFR can vary widely for some galaxies, particularly those that have experienced unusual star formation histories, are interacting or merging, or have AGN activity. Dusty infrared-luminous galaxies in general can have a wide range of SFR estimates which depend on the tracer that is used.  
For example, published SFR estimates for NGC\,1614 range from 27.4~M$_{\odot}$yr$^{-1}$ \citep{Adamo20} to 74.7~M$_{\odot}$yr$^{-1}$ \citep{Tateuchi15}, or a factor of 2.7, which would translate to a factor of 2.7 difference in $\Gamma$.  In this work, we assumed a SFR of 49.6~M$_{\odot}$yr$^{-1}$, which is $\approx$factor of two higher than that assumed by \cite{Adamo20}, and therefore decreased our estimated $\Gamma$ by a factor of two relative to theirs.  Another example of very different estimates of the SFR for a galaxy are found for the late-stage merger NGC~34, which has experienced strong changes in its star formation history over the past $\sim0.5$~Gyr (Zhang et al., in prep).   Published SFR estimates range between 5 and 90~M$_{\odot}$yr$^{-1}$, or a factor of 18 (!), with hydrogen recombination lines giving much lower estimates than infrared-based tracers.

Other assumptions that can potentially impact $\Gamma$ are related to the age-dating procedure itself. One key assumption, which we have explored in this work, is the maximum reddening value allowed during SED fitting to break the age-reddening degeneracy. 
We showed in $\S$ \ref{sec:agedating} that many clusters which are older than 10~Myr can be incorrectly dated to younger than 10~Myr when the maximum E(B-V) value allowed in the fit is too high.  This age-dating problem has been identified in a number of other studies as well (e.g., \cite{Whitmore23b,Chandar23b}; Thilker et al. submitted).
Age-dating issues can also affect cluster mass estimates.  For example, the most massive cluster younger than 6~Myr (H$\alpha$ bright) in NGC 1614 has an estimated mass that is 5 times higher in 
\cite{Adamo20} (who found M$=1\times10^8~M_{\odot}$) than that found here ($\sim2\times10^7~M_{\odot}$); their mass for this single cluster is 1.5 times more than the total mass we find for all $1-10$~Myr clusters combined.
Some reasons that cluster mass estimates might differ between different works are: (1)  the $\mathrm{M}/\mathrm{L}_\mathrm{V}$ changes by a factor of $\sim$2.5 for clusters with ages between 1 and 6~Myr (those predicted to have H$\alpha$ emission), which means that mass estimates for young clusters can vary by a similar factor; (2) clusters older than 10~Myr that are incorrectly fit to an age $<10$~Myr and moderate-to-high reddening will have artificially high mass estimates and be incorrectly included in the calculation for $\Gamma_{\rm 1-10~Myr}$.

We conclude that $\Gamma$ is sensitive to details of the assumptions made to calculate it.  
Galaxies with very high rates of star formation and $\Sigma_{\rm SFR}$ in the nearby universe often have rapidly changing star formation histories, AGN activity, and a significant amount of dust, making calculations of $\Gamma$ in these systems particularly challenging.

% Don't change these lines
%\bsp	% typesetting comment
\label{lastpage}
\end{document}